# Nonlinear elasticity of the extracellular matrix fibers facilitates efficient inter-cellular mechanical communication

Ran S Sopher, Hanan Tokash, Sari Natan, Mirit Sharabi, Ortal Shelah,
Oren Tchaicheeyan, Ayelet Lesman*

School of Mechanical Engineering, Faculty of Engineering, Tel-Aviv University, Israel

* Corresponding author:

Dr. Ayelet Lesman

Wolfson Mechanical Engineering Building, Room 331

School of Mechanical Engineering, Faculty of Engineering

Tel-Aviv University

Tel Aviv 69978

Israel

Tel: +972-3-6408233

Email: ayeletlesman@tauex.tau.ac.il

# Abstract

Biological cells embedded in fibrous matrices have been observed to form inter-cellular bands of dense and aligned fibers, through which they mechanically interact over long distances. Such matrix-mediated cellular interactions have been shown to regulate a variety of biological processes. The current study was aimed at exploring the effects of elastic nonlinearity of the fibers contained in the extracellular matrix (ECM) on the transmission of mechanical loads between contracting cells via the ECM. Based on our biological experiments with fibroblasts in fibrin gels shortly after seeding, we developed a finite-element model of two contracting cells embedded within a fibrous network. The individual fibers were modeled as showing either linear elasticity, compression-microbuckling, tension-stiffening or both of the latter. Fiber compression-buckling resulted in smaller loads occurring in the ECM, but these were more directed toward the neighboring cell. The latter decreased with increasing cell-to-cell distance; when cells were >15 cell-diameters apart, no such inter-cellular interaction was observed. Tension-stiffening further contributed to directing the loads toward the neighboring cell, though to a smaller extent. The contraction of two neighboring cells resulted in mutual attraction forces, which were considerably increased by tension-stiffening, and decayed with increasing cell-to-cell distances. Nonlinear elasticity contributed also to the onset of force polarity on the cell boundary, manifested in larger contractile forces on the part of the cell boundary pointing toward the neighboring cell. The density and alignment of the fibers within the inter-cellular band were considerably greater when fibers buckled under compression, with tension-stiffening further contributing to this structural remodeling. While previous studies have established the role of the ECM nonlinear mechanical behavior in increasing the range of inter-cellular force transmission, our model demonstrates the contribution of nonlinear elasticity of biological gels to directionality and efficiency of mechanical-signal transfer between distant cells.

# 1. Introduction

The cellular actomyosin machinery actively generates forces that are transmitted to the cell surroundings to induce loads (displacements, strains and stresses) within the extracellular matrix (ECM); these can persist hundreds of microns away (1), and are known to influence cell morphology, migration and differentiation (2). Long-range loads have been thus proposed as a means for cells to mechanically communicate with each other, and demonstrated to play a key role in various biological, physiological and pathological processes, as diverse as capillary sprouting (3), cancel invasion (4), heart-beat synchronization (5) and morphogenesis (6).

The fibrous ECM demonstrates nonlinear elastic behavior that is manifested in compressive-softening and tension-strain-stiffening (7, 8). These are attributed to the mechanical behavior of the individual fibers contained in the matrix, showing strain stiffening in tension (8–12) and microbuckling under compression (9, 13, 14). The fibrous structure of the ECM also contributes to its macroscale elastic nonlinearity as a result of fiber reorganization under applied loading, manifested in fiber realignment and densification (15–17). Previous experimental studies have demonstrated that the nonlinear elasticity of the matrix act as a mechanism facilitating long-range transmission of loads, enabling cells to sense and respond to mechanical signals sent by other cells located at far distances. For example, Notbohm *et al.* (18) found that the contraction of fibroblasts within a fibrin matrix of nonlinear elastic behavior induced displacements that travelled considerably further than predicted when assuming linear elasticity. Vanni *et al.* (19) similarly found that the contraction of a single fibroblast within collagen gel induces strains that can propagate up to 800 μm through the substrate. Winer *et al.* (20) found that local strain-stiffening of fibrin gels facilitates the transmission of forces between fibroblasts or mesenchymal stem cells up to ~500-μm (~30-cell-diameters) apart. This is in a striking contrast to cells cultured on linear elastic gels, which sense and respond to loads induced by other cells in a distance limited to ~25 μm (21). These experimental findings have been elucidated by numerous analytical procedures, finite-element (FE) simulations and other computational models developed to study the effects of ECM elastic nonlinearity on the transmission of cell-contraction-induced loads through the network in which the cell(s) are embedded. Safran and colleagues (22–24) presented an analytical model demonstrating the long-range decay of displacements induced by the contraction of a single circular cell embedded in a medium showing nonlinear-elastic behavior (manifested in tension-stiffening and compression-softening). These predications were later supported by computer simulations in fibrous networks demonstrating that fiber buckling results in displacements and stresses travelling considerably farther than they would in a linear-elastic medium (25, 26). Also important, the fibrous structure of the ECM, and particularly the load-induced geometric rearrangement of the network, have been found to contribute to the nonlinear elastic behavior (especially stiffening) of the ECM and to act as a mechanism considerably increasing the range of inter-cellular force transmission and sensing (1, 15–17, 27–30). These studies distinguish the ECM nonlinear mechanical behavior, and particularly fiber microbuckling, as a mechanism supporting long-range force transmission through the ECM. Less is known about the effect of the mechanical behavior of the fibers constituting the ECM on force transmission between cells.

Previous experimental studies (3, 4, 27, 31, 32) have revealed that contraction of multiple cells embedded in fibrous biological gels induces structural remodeling of the fibers in the inter-cellular medium manifested in the formation of aligned and densely packed fiber 'bands' connecting the contracting cells. This observation demonstrates not only the long-range nature of the force transmission through the ECM – which is both facilitated by and contributing to such structural remodeling – but also reflects the tendency of cell-contraction-induced loads to be delivered in a highly directional manner toward neighboring cells. Harris *et al.* (31) was the first to show that the forces exerted by tissue explants placed millimeters apart in collagen gel stretched and aligned the fibers between the explants, thereby forming inter-cellular 'bands'. More recent studies have demonstrated the ability of cells to respond to such inter-cellular bands. Korff and Augustin (3) found that forces applied by two endothelial-cell spheroids embedded in collagen gel and placed up to 700 μm apart, resulted in directional sprouting of capillaries along aligned fibrils between the spheroids. Shi *et al.* (4) further showed that the directional remodeling of collagen fibers accelerated the transition of gel-cultured mammary-acini cells to an invasive phenotype. Such biological observations highlight the importance of the structural remodeling of the inter-cellular matrix in supporting long-range inter-cellular mechanical interactions, thereby regulating various biological processes. Still, the physical mechanisms facilitating the formation of such ECM 'bands' and the manner in which they mediate force transmission between cells are poorly understood. Particularly, despite the fact that several computational models were able to capture the tendency of loads to concentrate within the inter-cellular medium and align the fibers contained in this region (16, 27, 29, 30, 32), a quantitative exploration of the influence of ECM elastic nonlinearity on regulating matrix-mediated mechanical interaction between cells is warranted.

In this work, we explore the contribution of the nonlinear elastic properties of the ECM fibers to the structural remodeling of the inter-cellular ECM and to the transfer of mechanical loads between neighboring cells. Experimentally, we show the ability of fibroblast cells to structurally align and densify the fibers between them even shortly after seeding in fibrin gels, when they are still primarily spherical. Based on our experimental setting, we developed FE simulations of two contracting cells (separated by 1.5-19 cell diameters) embedded within fibrous nonlinear elastic networks. The simulation outcomes indicate that cell-contraction-induced loads are highly directed toward neighboring cells owing to the nonlinear elastic behavior of the matrix and its constituting fibers; this observation is coupled with elevated structural remodeling of the inter-cellular region of the ECM. We link these observations with efficient transfer of mechanical loads between cells. We also show that inter-cellular interactions manifest in attraction forces occurring between neighboring cells, and also lead to the onset of force polarity on the cell boundaries. The model presented herein contributes to the understanding of biological processes involving ECM-mediated interactions that can influence cell differentiation, migration, and morphogenesis; it can also expand the knowledge basis required for designing biomaterials that support efficient inter-cellular mechanical interactions.

## 2. Methods

### 2.1. Biological experiments

Approximately 5,000 NIH3T3 - GFP-Actin cells were seeded in 20 μl of fibrin gel (5 mg/ml fibrinogen) labelled with Alexa Fluor 456 as described in previous studies (e.g. (18)). The gel was scanned at several time points post seeding using confocal laser-scanning microscope (Zeiss LSM 880 lens x40, water immersion) to capture distant cells forming 'bands' between them, as visual evidence of mechanical interaction (18). Image analysis was followed using ImageJ (NIH, Bethesda, MD, USA; https://imagej.nih.gov/ij/) and the OrientationJ plugin (EPFL, Switzerland, 2017; http://bigwww.epfl.ch/demo/orientation/#soft) to examine the orientation and intensity (indicative of the fibrin density, and normalized to the mean intensity of the entire image) of several regions contained in the ECM between two interacting cells or far away from the cells.

### 2.2. Computational modeling

In order to explore the effects of the mechanical behavior of the ECM on the mechanical loads occurring within the medium between contracting cells, a FE model was developed. This simulated two identical contracting cells embedded in a fibrous network, while the cell-to-cell distance and the mechanical properties of the fibers contained in the network were adjusted to create 558 model variants (31 cell-to-cell distances, 6 types of fiber mechanical properties, 3 levels of cell contraction). The model was greatly based on computational models described in previous studies (mainly Notbohm *et al.* (26) and Liang *et al.* (33)) with some major modifications, which are described below.

*2.2.1. Network geometry and architecture*

A two-dimensional (2D) array of multiple identical square box-X units was initially created. Each unit contained four corners acting as nodes and two horizontal, two vertical and two diagonal sides acting as elements. Each of the four horizontal/vertical sides was shared with another X-box unit; similarly, each of the four corners was shared between four identical units meeting at this corner, resulting in each corner/node being the edge of eight sides/elements (referred to by Notbohm *et al.* (26) and others as connectivity=8; Figure 1a). The locations of all nodal positions were modified by relocating each node to a randomly selected location contained within a circular region of radius equal to the length of the horizontal/vertical sides of the aforementioned square box-X. This created an array of elements of different lengths, where each linear element connecting a pair of nodes represented a fiber segment stretched between two cross-linking points, thereby simulating the fibrous network contained in the ECM. Elements were removed from the network to create two circular void regions (15, 16, 25, 26, 33) representing cells embedded within the ECM; the centers of both circles were coincident with a horizontal line passing through the center of the network (Figure 1b). Equivalent model variants with a single cell each (the center of which was coincident with the center of the network) were also generated. Cell diameter and mean fiber-segment length were assumed to be 15.2 μm and 3.5 μm, respectively, according to our biological experiments with fibroblast cells embedded in 5 mg/ml fibrin gel (Figure 2). Cell-to-cell distance ranged between 1.5 to 19 cell diameters (*i.e.*, approximately 23-289 μm, based on our laboratory experiments; full list of cell-to-cell distances: 1.5, 1.8, 2.1, 2.9, 3.4, 4.0, 4.6, 5.3, 5.9, 6.5, 7.1, 7.8, 8.4, 9.0, 9.6, 10.3, 10.9, 11.5, 12.1, 12.8, 13.4, 14.0, 14.6, 15.2, 15.9, 16.5, 17.1, 17.8, 18.4, 19.0). The diameter of the network was 100 times larger than that of the cell (*i.e.*, 1520 μm), so that the distance between each cell and the boundaries of the network was at least twice larger than the cell-to-cell distance.

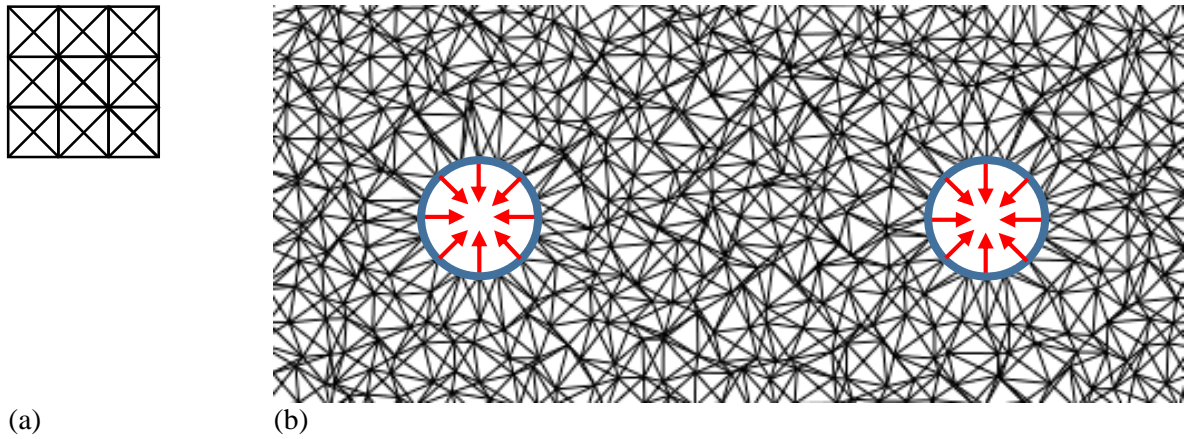

(a)  (b)

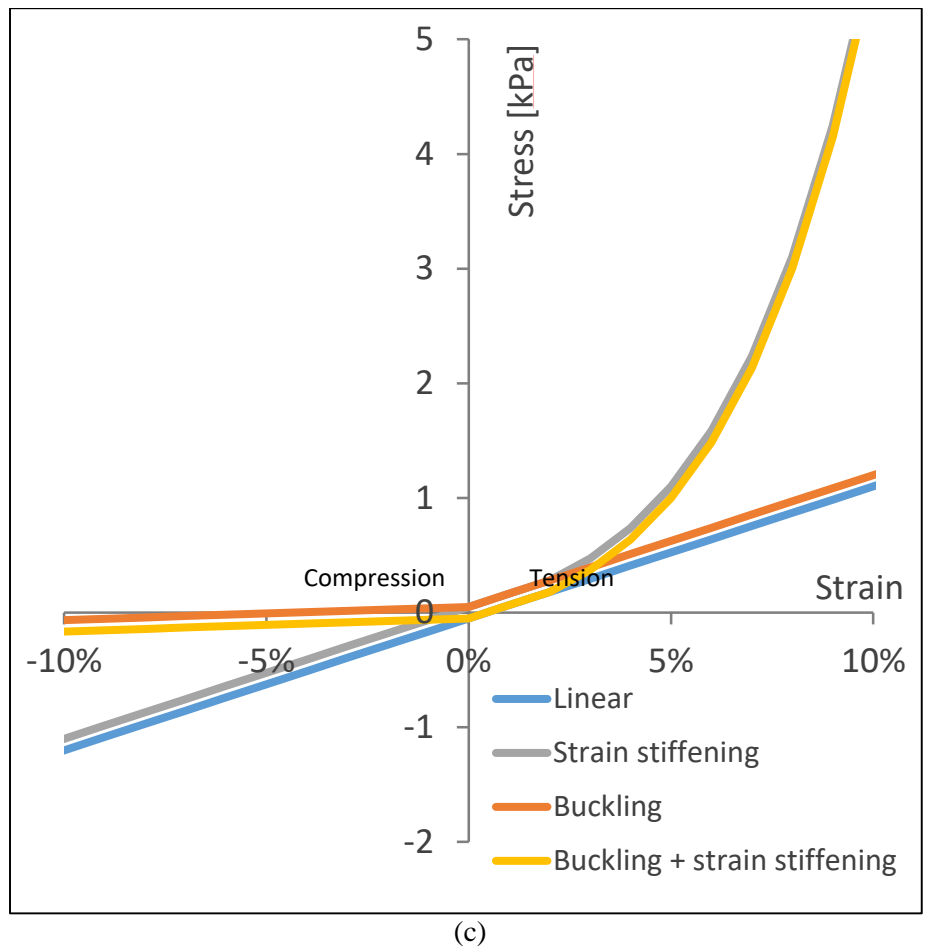

(c)

**Figure 1:** Details of the finite-element (FE) model: (a) The elementary unit of the model (before randomization). Nine such units are shown. (b) A small portion of a two-dimensional FE model of two cells contracting within the fibrous network. The contracting cells are modeled as void regions (blue circles) to which a boundary condition of radial contractile displacement (red arrows) is applied. (c) Stress-strain curves representing the mechanical properties assigned to each of the fiber elements consisting the FE model. Four material models were used to simulate the mechanical behavior of the ECM fibers, as listed in Section 2.2.2. Lines are slightly shifted for visualization purposes.

### 2.2.2. Mechanical properties

Four material models were used to simulate the mechanical behavior of the individual ECM fibers (Equation 1, Figure 1c; (26, 33)): i) 'linear': linear elastic material with tensile and compressive elastic moduli (Young's moduli, E) of 11.5 kPa ($\rho=1$); ii) 'buckling': elastic material with tensile E of 11.5 kPa and compressive E 2- (buckling $^{1/2}$, ρ=0.5), 5- (buckling $^{1/5}$, ρ=0.2) or 10- (buckling $^{1/10}$, ρ=0.1) times smaller, which simulates fiber buckling;  iii) 'strain stiffening': hyper-elastic material with compressive E of 11.5 kPa and tensile E of 11.5 kPa within the logarithmic-strain (*i.e.* true strain) range of 0-2% (ρ=1); for tensile strains larger than 2%, stress increases exponentially to simulate strain-stiffening behavior; iv) 'buckling+stiffening': hyper-elastic material with tensile E of 11.5 kPa within the logarithmic-strain range of 0-2%, which increases exponentially for strains larger than 2%; compressive E is 10 times smaller than 11.5 kPa (ρ=0.1). Further information about the assignment of mechanical properties to the ECM fibers is included in the Supplementary Materials.

$$E = \begin{cases} \rho \cdot E_{ref}, & \lambda < 0 \\ E_{ref}, & 0 \leq \lambda < \lambda_s \\ E_{ref} \cdot e^{\frac{\lambda - \lambda_s}{\lambda_0}}, & \lambda \geq \lambda_s \end{cases}$$

**Equation 1**: Values of elastic modulus (Young's modulus, E) assigned to the truss elements of the FE model. $E_{ref}$=11.5 kPa is the reference elastic modulus, λ=Δ*l*/*l* is the engineering strain occurring within an element of length *l*, $\lambda_s$=2% is the strain above which strain stiffening occurs, $\lambda_0$=0.05 is the strain stiffening coefficient and ρ is the buckling ratio (9, 33).

### 2.2.3. Boundary conditions

Cell contraction was modeled by applying a boundary condition of radial contractile displacement (equals to 10%, 25% and 50% of the cell radius) to all nodes constituting the cell boundaries (Figure 1b) (in line with (26–30, 33)). The circular boundary of the entire network was fixed for translations and rotations in all directions (15, 29, 30, 33). In all simulations, the possibility of the boundary fixation affecting the model outcome measures was eliminated by ensuring that the strains, stresses and strain energy densities (SEDs) occurring along the network boundaries were negligible compared with those occurring in the cell vicinity (15).

### 2.2.4. Numerical method

Linear truss elements (*i.e.* supporting uniaxial tension and compression and unrestrained rotation about the nodes, which represent fiber cross-links, but with infinite resistance to bending) were used to model the fiber segments, following our preliminary study showing marginal differences in the model outcomes between truss- and beam- element networks (in line with (15, 16, 33)). The cross-sectional area of all elements was assumed to be constant at 0.031416 μm² (see Supplementary Materials). The model included approximately 590,000 elements and 148,000 nodes. The ABAQUS® Standard/Implicit FE solver (ABAQUS, Version 2017, Dassault Syst`emes Simulia Corp., Providence, RI, 2017) in its nonlinear analysis mode was used to process all model variants, with running time of approximately 5 minutes per simulation (on an i7-, 3.60GHz- CPU and 32 GB-RAM station).

*2.2.5. Outcome measures*

Values of displacement, strain, stress (logarithmic/true tensile and compressive strains and stresses), SED and reaction force were calculated at the nodes and/or centroids of all elements. A script coded in MATLAB® (version R2016b, MathWorks Inc., Natick, MA, USA) was used to determine the following outcome measures for all model variants: (a) Total contraction force: the sum of all reaction forces occurring in the nodes constituting the cell perimeter and pulling them along the radial axis toward the cell center (arrows at the top panel of Figure 7a). (b) Net cell-interaction force: the sum of projections of the radial contraction forces on the line connecting the centers of the two cells. This outcome measure is indicative of the level of inter-cellular attraction (negative) or repulsion (positive). (c) Contraction-force front-to-rear polarity ratio: the fraction of the total contraction force occurring within a 60° arc of the cell boundary pointing toward the neighboring cell (orange arc at the top panel of Figure 7b) to that occurring in the arc pointing toward the exactly opposite direction (sky-blue arc). This outcome measure is used to evaluate the relative amount of cell-contraction force directed into sending signals to the neighboring cell; it can also be used as an indication of the direction (toward/away from the neighboring cell) where the cell is likely to spread or migrate (force polarity is known to affect cell motility and morphology; (34)). (d) The mean loads (strains, stresses and SED) occurring within a disc surrounding an individual cell, of radius equals to half of the cell-to-cell distance (orange disc at the top panel of Figure S6a); (e) Directionality ratio: the fraction of the sum of loads occurring in all elements falling within a 60° sector pointing toward the neighboring cell (orange sector at the top panel of Figure 5a) to the sum of loads occurring in all elements falling within the entire aforementioned disc. (f) Asymmetry ratio: the fraction of the sum of loads occurring in all elements falling within the aforementioned 60° sector to the sum of loads occurring in all elements falling within a 60° sector pointing toward the opposite direction (sky-blue sector at the top panel of Figure 5b; similar to the 'signal parameter' described elsewhere (15)). The two latter outcome measures are used to evaluate the relative amount of the load (strain, stress or energy) caused by cell contraction directed toward the neighboring cell (in very simplistic words, how much of the effort that the cell is putting while contracting is potentially delivered to the neighboring cell, or how efficient the cell contraction is in terms of inter-cellular mechanical signal transfer); directionality and polarity/asymmetry ratios of 0.17 and 1.0, respectively, indicate no preferred orientation of loads toward the neighboring cell. (g) The relative change (in %) in density of elements contained in the 60° sector pointing toward the neighboring cell (orange sector at the top panel of Figure 8a) as a result of cell contraction; (h) Change in the mean angle of the orientation of the fibers contained in aforementioned 60° sector as a result of cell contraction. The latter two outcome measures were calculated in view of previous experimental studies showing that contraction of neighboring cells causes densification and realignment of the ECM fibers stretched between the cells; cells may respond to such structural changes, for example by migrating along the aligned fibers (34). All aforementioned outcomes were derived for both cells contained in each model variant; the mean of the outcomes of the two cells was then calculated, and is regarded as the model outcome (as presented below).

## 3. Results

### 3.1. Biological experiments of inter-cellular mechanical interactions

Actin-GFP fibroblast cells were embedded in fluorescently labelled fibrin gels at a low cellular density such that cells were well separated from each other. Within two hours from cell seeding, while most of the cells were still of a spherical shape, cells were observed to deform the fibrous matrix, creating highly remodeled matrix 'bands' between neighboring cell pairs (Figure 2a). Image analysis of the fibrous structure of the ECM revealed elevated fiber density and alignment within the inter-cellular medium compared with other regions of the ECM (Figure 2b,c). This observation demonstrates the ability of cells, still in their spherical shape, to deform the surrounding matrix in a highly directional manner toward neighboring cells. We used these biological experiments as the basis of a computational study aimed at examining the contribution of the nonlinear elastic behavior of the fibrous matrix to mechanical interaction between cells. Our model was designed to test the experimentally-observed directionality of deformations occurring in the ECM, when isolated from other factors potentially creating load anisotropy (*e.g.* elongated cell geometry or non-uniform contraction). We therefore developed a FE model of two uniformly contracting circular cells situated at increasing distances from each other within a structural fibrous mesh, and explored the effects of fiber elastic nonlinearity on the distribution and propagation of loads (forces, displacements, strains, stresses and SEDs) through the matrix, and particularly in the region between neighboring cells.

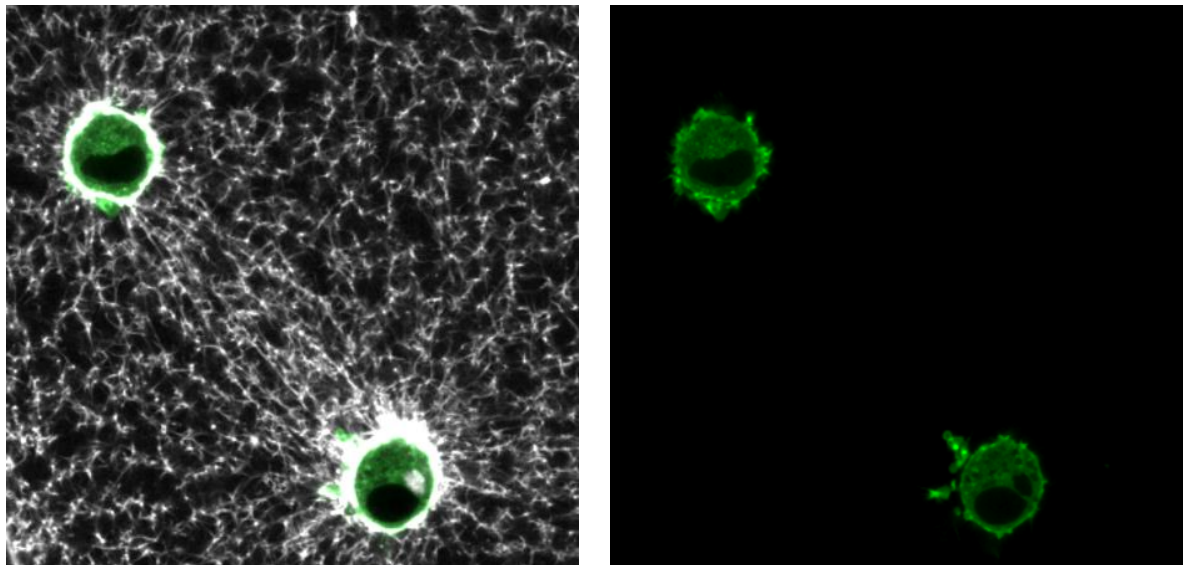

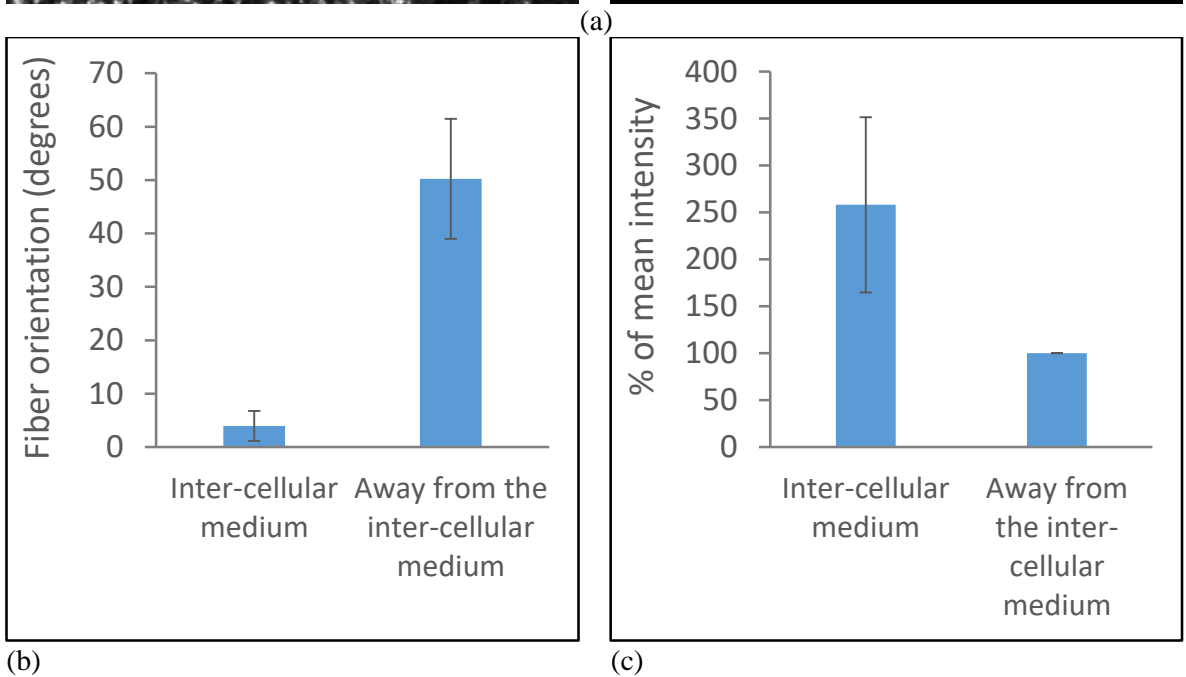

**Figure 2:** (a) Two NIH3T3 - GFP-Actin cells (green) embedded in fibrin gel labelled with Alexa Fluor 456 (grey). The images show the overlay of cells and fibrin matrix (left) and only-cells image (right). (b) Orientation of the fibrin fibers contained in the ECM in the inter-cellular matrix and away from the cells; 0° is defined as the line connecting the centers of the two cells. (c) Mean intensity as measured both within the inter-cellular medium and away from the cells. Data in (b) and (c) are derived from six pairs of cells as demonstrated in (a).

### 3.2. Finite-element analysis of the transmission of loads induced by cell contraction through a fibrous matrix

Our FE simulations of a single cell contracting within a fibrous matrix show that the distribution of loads within the matrix was not homogeneous. Particularly, the majority of loads were carried through a small number of fiber segments constituting 'force chains' (mostly apparent in the buckling model variants; Figure 3 and Figure S3). Despite this non-homogeneity, the load distribution around a single cell was nearly isotropic about the cell center and did not demonstrate any preferred orientation. Tensile loads were most elevated in the fibers aligned approximately perpendicularly to the cell edges (thereby forming 'tethers' or 'force chains' propagating away from the contracting cell; Figure 3a,b and Figure S3a-d). The compressive strains, which were generally of magnitude approximately twice-larger than the tensile strains, were the largest within the fibers aligned approximately tangentially to the cell perimeter (thus forming 'rings' around the cell; Figure 3c,d and Figure S3e-h). The model variants simulating the contraction of a pair of neighboring cells similarly demonstrated propagation of loads through distinct fibrillar paths, with tensile-strain tethers forming between the cells and compressive-strain rings encircling the two cells (Figure 4a-d and Figure S4a-h). These observations are attributable to the contraction of the cell pulling mainly the fibers perpendicular to the cell edges and squeezing primarily the tangential ones. The tendency of contraction-induced loads to concentrate in such tethers and rings was greatly augmented by fiber compression-buckling, with fiber tension-stiffening marginally contributing to this effect (Figure S3 and Figure S4). Fiber compression-buckling expectedly resulted in the compressive strains occurring in the ECM as a result of cell contraction being larger, as opposed to the tensile loads and SEDs being considerably smaller; tension-stiffening again only marginally affected the magnitude of loads (Figure S3 and Figure S4). It can be therefore concluded that the contraction of isolated single cells induces substrate deformations that are symmetrically distributed about the cell center, yet propagate through distinct fibrillar arrangements, with magnitudes that are predominantly dictated by fiber microbuckling; the propagation of loads through tethers or rings and their dependency on fiber buckling is also manifested in the contraction of two neighboring cells.

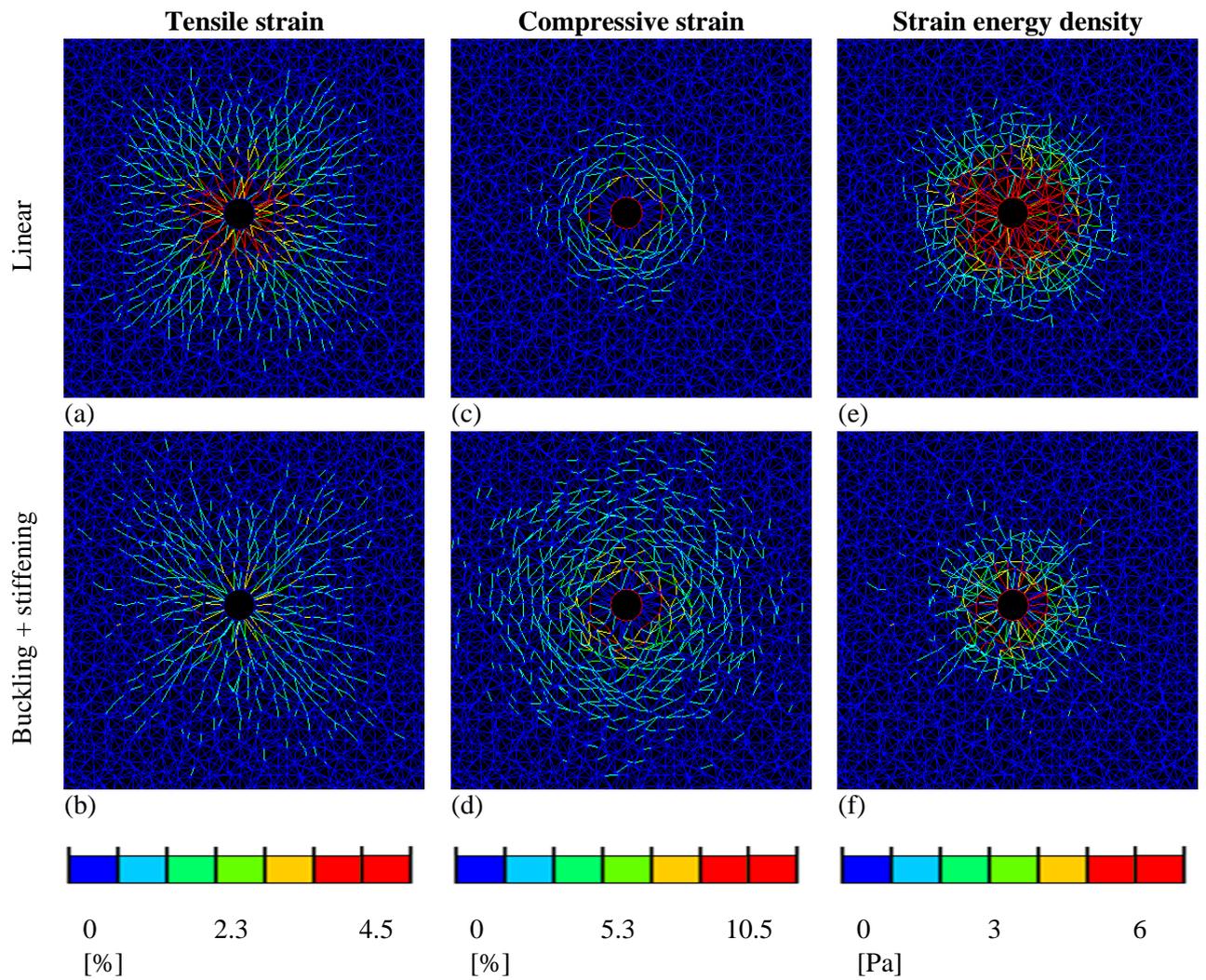

**Figure 3:** Contour plots showing the tensile (logarithmic) strains (left column), compressive (logarithmic) strains (middle column) and strain energy densities (SEDs, right column) occurring in the fiber segments within the vicinity of a single, isolated contracting cell, for 25% contraction. Plots were produced for two of the material models used to simulate the mechanical behavior of the individual fibers (linear, buckling + strain stiffening; Figure 1c). Equivalent plots produced for all four material models used to simulate the mechanical behavior of the individual fibers (Figure 1c) are contained as supplementary materials.

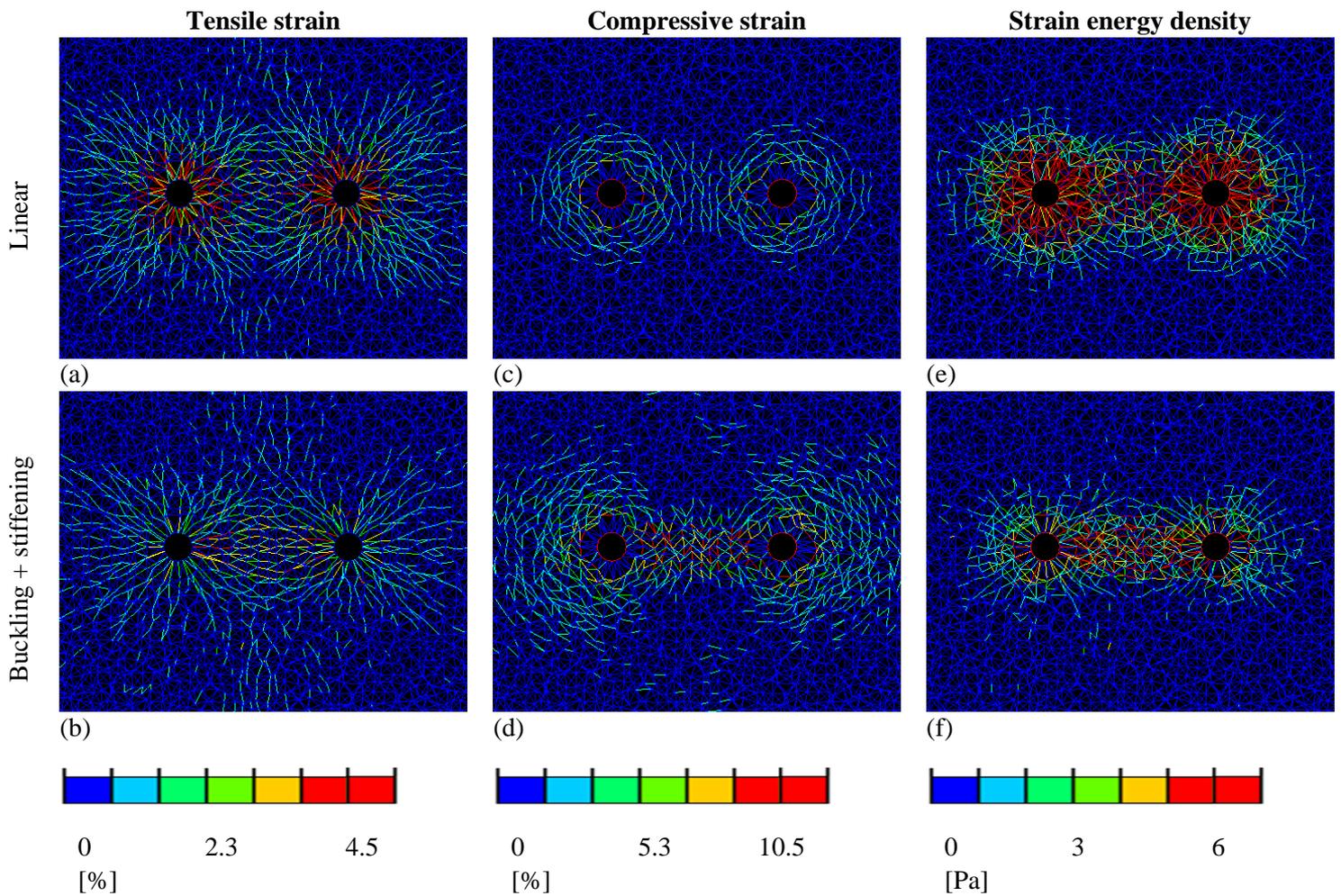

**Figure 4:** Contour plots showing the tensile strains (left column), compressive strains (middle column) and SEDs (right column) occurring in the fiber segments within the vicinity of two neighboring contracting cells (here the cell-to-cell distance is equivalent to 3.4 cell diameters as an example), and particularly within the inter-cellular medium, for 25% contraction. Plots were produced for two of the material models used to simulate the mechanical behavior of the individual fibers (linear, buckling + strain stiffening; Figure 1c). Equivalent plots produced for all four material models used to simulate the mechanical behavior of the individual fibers (Figure 1c) are contained as supplementary materials.

### 3.3. The effect of fiber compression-buckling on load transfer between cells

We next analyzed the distribution of loads occurring in the vicinity of two neighboring contracting cells in more quantitative terms. The contour plots in Figure 4 and Figure S4 show that fiber microbuckling resulted in the tensile strains and SEDs occurring within the cell vicinity being generally smaller, yet more concentrated within the inter-cellular medium; fiber linear elasticity produced more even distribution of loads around each cell. The mean tensile strain and SED occurring in a disk surrounding each of the cells revealed considerably lower values for the compression-buckling model variants compared with those where fibers resisted compression (Figure S6a and Figure S6c). This indicates that it is 'easier' for cells to contract within a bucklable matrix, which is ascribable to its lower resistance. To quantify the tendency of loads to concentrate in the inter-cellular medium, which arises from the contraction of two cells practically stretching this region of the ECM, we calculated the fraction of deformations and SEDs falling within the matrix area pointing toward the neighboring cell (outcome measure e in Section 2.2.5, referred to as 'directionality ratio'). We further evaluated these loads against those falling within the ECM area pointing toward the opposite direction (outcome measure f in Section 2.2.5, referred to as 'asymmetry ratio'). We found that fiber buckling under compression resulted in the distribution of SEDs, tensile strains and compressive strains being more directed toward the neighboring cell (*i.e.*, concentrating in the inter-cellular region), as reflected in higher directionality and asymmetry ratios (for 10%, 25% and 50% cell contraction; Figure 5a,b, Figure S7, Figure S8 and Figure S9). The most distinguished effect was found for SEDs at the shortest cell-to-cell distances: at cell contraction of 25% and cell-to-cell distance of 2.1 cell diameters, for example, fiber buckling resulted in approximately 0.40 of the total SEDs being directed toward the neighboring cell, compared with 0.24 for linear-elastic fibers and 0.17 for a single, isolated contracting cell (Figure 5a). 'Stronger' buckling ratios (*i.e.* smaller $\rho$ in Equation 1) normally resulted in a more considerable fraction of loads falling within the inter-cellular medium (small panel at the top right of Figure 5a), which further emphasizes the critical role of fiber buckling in directing the ECM cell-contraction-induced loads toward the neighboring cell. Asymmetry ratios, which reflect the asymmetric distribution of the loads about the cell center, were generally more affected by fiber buckling than the directionality ratios. In particular, in the compression-buckling model variant, the total SED falling within the fraction of matrix pointing toward the neighboring cell was approximately 2.8-fold larger than on the opposite side (*i.e.* asymmetry ratio of 2.8), as compared with only 1.4 in the equivalent linear-elastic model variant, and 1.0 for a single, isolated cell (example values are for cell contraction of 25% and cell-to-cell distance of 2.1 cell diameters, Figure 5c). The effect of microbuckling was noticeable up to a distance of 9 cell diameters, above which the directionality and asymmetry ratios approached the values for a single, isolated cell (0.17 and 1.0, respectively, which indicate no preferred orientation of loads toward the neighboring cell), implying that mechanical inter-cellular signaling no longer occurred (Figure 5a,b, Figure S7, Figure S8 and Figure S9). In general, loads were smaller and all the above-mentioned trends were weaker for smaller cell contraction (25% versus 10% cell contraction; Figure S7 and Figure S8). Overall, the results indicate that fiber microbuckling results in lower-magnitude loads that are more efficiently directed to the neighboring cell.

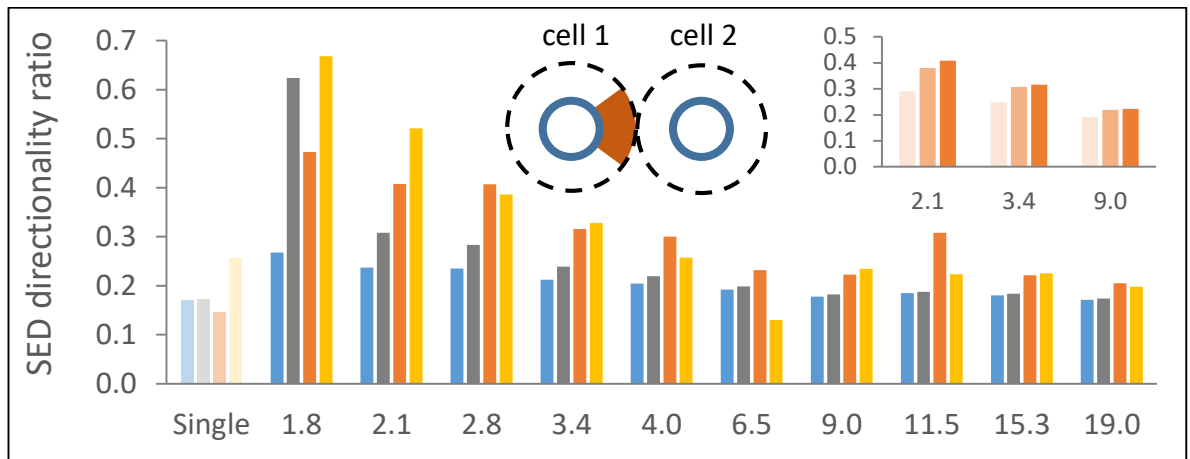

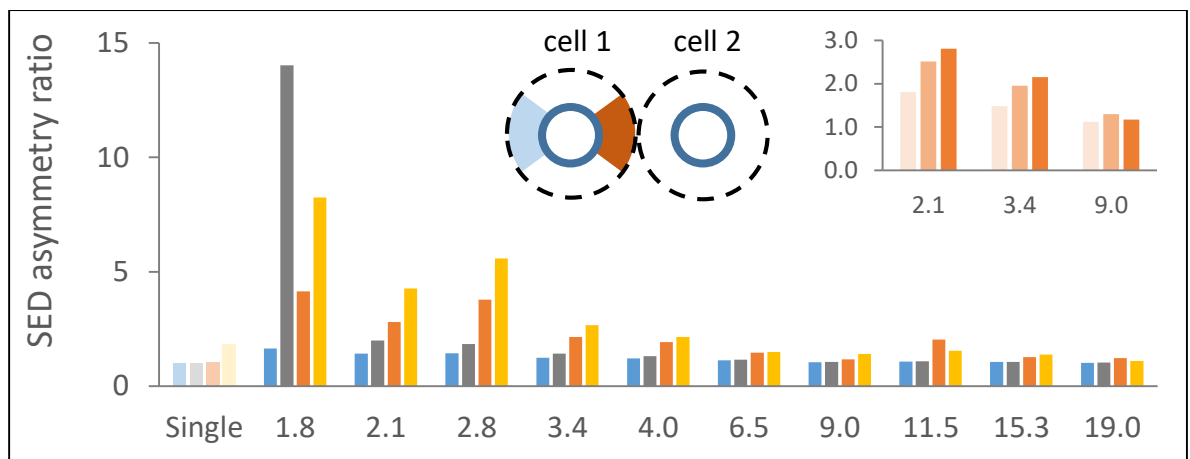

**Figure 5:** Directionality (a) and asymmetry (b) ratios of SED. The model variants shown include several cell-to-cell distances (in terms of cell diameter, D), and four of the material models used to simulate the mechanical behavior of the individual fibers (Figure 1c), for 25% cell contraction. Directionality and asymmetry ratios of 0.17 and 1.0, respectively, indicate no preferential orientation of loads toward the neighboring cell. The values calculated for the single-cell model variants (assuming a disc of radius equals to half of the cell-to-cell distance of 2 D) are shown for comparison. On the top right, an equivalent bar chart with fewer cell-to-cell distances and three material models used to simulate the mechanical behavior of the individual fibers, representing different buckling ratios ($\rho=1/2$, $\rho=1/5$ and $\rho=1/10$).

### 3.4. Effect of fiber tension-strain-stiffening on load transfer between cells

Fiber strain-stiffening in tension further contributed to concentration of loads in the inter-cellular medium as reflected by higher directionality and asymmetry ratios (Figure 5b,c, Figure S7, Figure S8 and Figure S9). The effect of stiffening was most evident in smaller cell-to-cell distances and larger cell contractions (50% as opposed to 25% and 10%), which is attributable to the tensile strains exceeding the critical stiffening threshold of 2% tensile strain (Equation 1). For cell contraction of 25% and cell-to-cell distance of 1.8 cell diameters, fiber stiffening resulted in directionality ratio of 0.62 and asymmetry ratio of 14, compared with 0.27 and 1.6 for the equivalent linear-elastic model variant, and 0.17 and 1.0, respectively for a single, isolated cell (Figure 5b,c). At that short distance, the matrix between cells experiences high level of tensile strain, thereby resulting in higher SEDs (cells need to stretch more to achieve a certain level of contraction; discussed in Section 4). When fibers were modeled as both buckling under compression and stiffening in tension, the directionality and asymmetry ratios were substantially elevated, reaching values of up to 0.52 and 4.3, respectively, at cell contraction of 25% and cell-to-cell distance of 2.1 cell diameters (Figure 5a,b). For 50% cell contraction, fiber nonlinear elasticity contributed to directionality ratio of up to 0.80 at the cell-to-cell distance of 2.1 cell diameters (Figure S9). For cell contraction of 10%, the differences between equivalent strain-stiffening and non-strain-stiffening model variants were marginal (Figure S7), which is ascribable to the tensile strains rarely exceeding 2.5% (strain stiffening was assumed to occur above 2%; Section 2.2.2). The influence of fiber tension-stiffening was noticeable up to approximately 3-4 cell diameters, a relatively smaller distance than that found for the effect of compression-buckling (9 cell-diameters, Section 3.3). Overall, in larger cell contractions and smaller cell-to-cell distances, fiber strain stiffening in tension – alone or in combination with compression microbuckling – further contributes to concentration of loads within the inter-cellular region of the matrix, as demonstrated in the amplification of the directionality and asymmetry ratios of the load distributions.

### 3.5. The effect of fiber nonlinear elasticity on cell-induced forces

Analysis of the force balance acting on the cell boundaries (Figure 6 and Figure S10) and its dependency on the cell-to-cell distance can serve as an indication of how the ECM mediates the propagation of forces between contractile cells. A key question is whether the contraction of two neighboring cells results in mutual attraction or repulsion. We thus calculated the projection of the net force pulling the cell boundary toward the cell center (contraction force) on the line connecting the centers of the two cells (outcome measure b in Section 2.2.5, referred to as 'net cell-interaction force') and plotted it against the cell-to-cell distance (Figure 7a, Figure S11a and Figure S12a). We found that the direction of the net force that the cell is applying to the ECM is consistently the direction opposite the neighboring cell. It can thus be inferred that in our simulations the cells always attract each other. The cell-interaction force was only slightly influenced by fiber compression-buckling, but considerably affected by tension-stiffening (Figure 7a, Figure S11a and Figure S12a). For example, at cell contraction of 25% and cell-to-cell distance of 2.1 cell diameters, fiber stiffening resulted in the interaction force exceeding 0.4 nN, compared with 0.07 nN (*i.e.* 5.7-

fold) for the equivalent linear-elastic model variant (Figure 7a). This effect of strain stiffening was most evident for the larger levels of cell contraction; for 50% contraction, there was an almost 30-fold difference in the interaction force between the strain-stiffening and linear-elastic cases (Figure S12a). The cell-interaction force decreased with increasing cell-to-cell distance, and for all levels of cell contraction reached a plateau at distances larger than 6-7 cell diameters (Figure 7a, Figure S11a and Figure S12a).

Also of interest is the force polarity that develops on the cell boundary. This can potentially trigger cell morphological changes and migratory preference along the direction of force polarization (34). Therefore, we also computed the front-to-rear force polarity ratio (with 'front' referring to the direction pointing toward the neighboring cell; outcome measure c in Section 2.2.5), as indicative of the direction toward which the cell(s) are likely to spread or migrate (Figure 7b, Figure S11b and Figure S12b). We found that force polarity was greatly elevated as the cells were closer together, particularly when the matrix fibers exhibited both buckling and stiffening (Figure 7b). At cell contraction of 25% and cell-to-cell distance of 2.1 cell diameters, for example, the force applied in the direction pointing toward the neighboring cell was approximately 1.2- (linear-elastic fibers), 2.2- (compression-buckling), 2.0- (tension-stiffening) and 4.0- (buckling and stiffening) fold larger than the force applied in the opposite direction (Figure 7b). The force polarity considerably increased with the level of cell contraction; at 50% contraction this reached as much as 6.2 (tension-stiffening) and 22.7 (buckling and stiffening) (cell-to-cell distance of 2.1 cell diameters, Figure S12b); for a smaller cell-to-cell distance of 1.8 cell diameters, polarity exceeded 100. This is attributable to the large tensile deformations occurring in the fibers contained in the inter-cellular region of the matrix (Section 3.4). These findings indicate that nonlinear elasticity contributes to attraction forces and the onset of force polarity between neighboring contracting cell(s); cells are thus more likely to adapt their shape or migrate toward their neighbours when embedded within a nonlinear elastic matrix compared with linear-elastic ones.

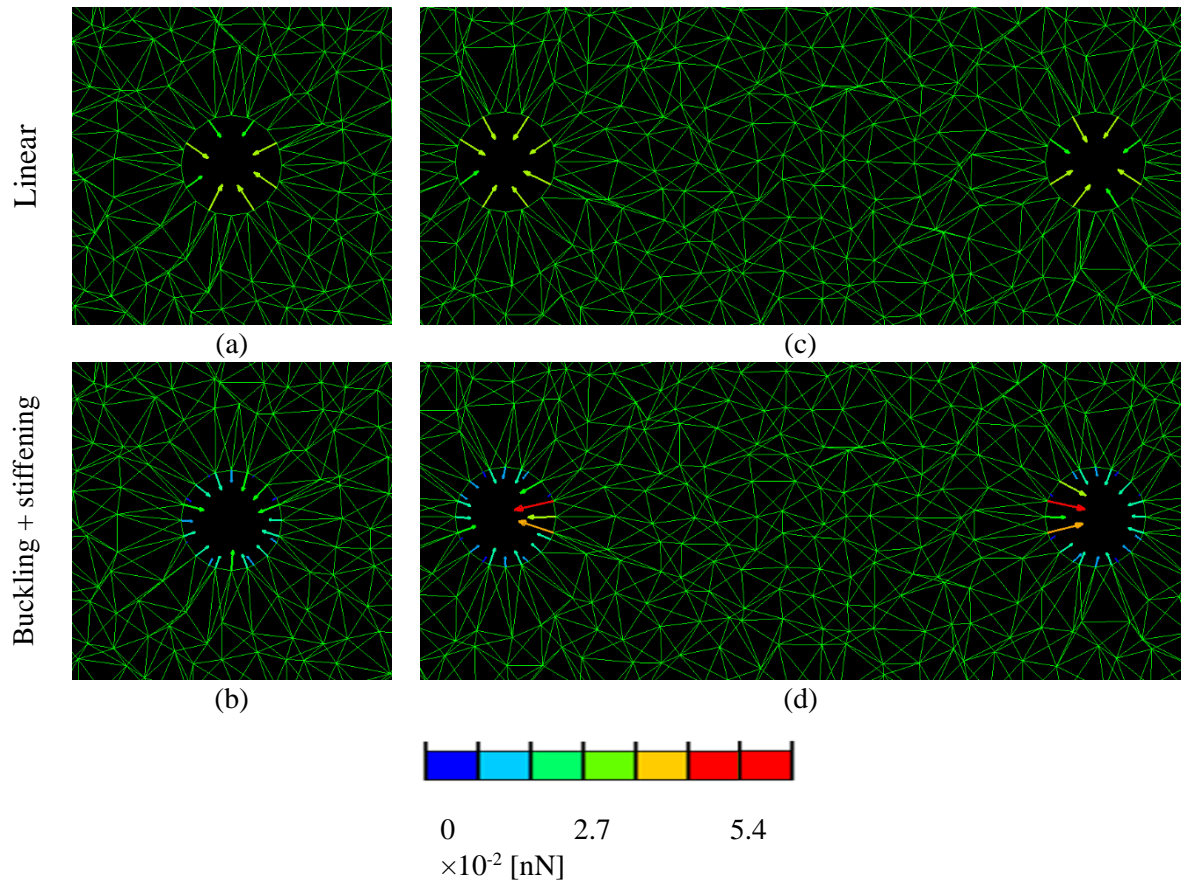

**Figure 6:** Contour plots showing the reaction forces occurring on the cell boundaries for a single (left column) or two (right column; here distance between the neighboring cells is equivalent to 3.4 cell diameters as an example) contracting cells, for 25% contraction. Plots were produced for two of the material models used to simulate the mechanical behavior of the individual fibers (linear, buckling + strain stiffening; Figure 1c). Equivalent plots produced for all four material models used to simulate the mechanical behavior of the individual fibers (Figure 1c) are contained as supplementary materials.

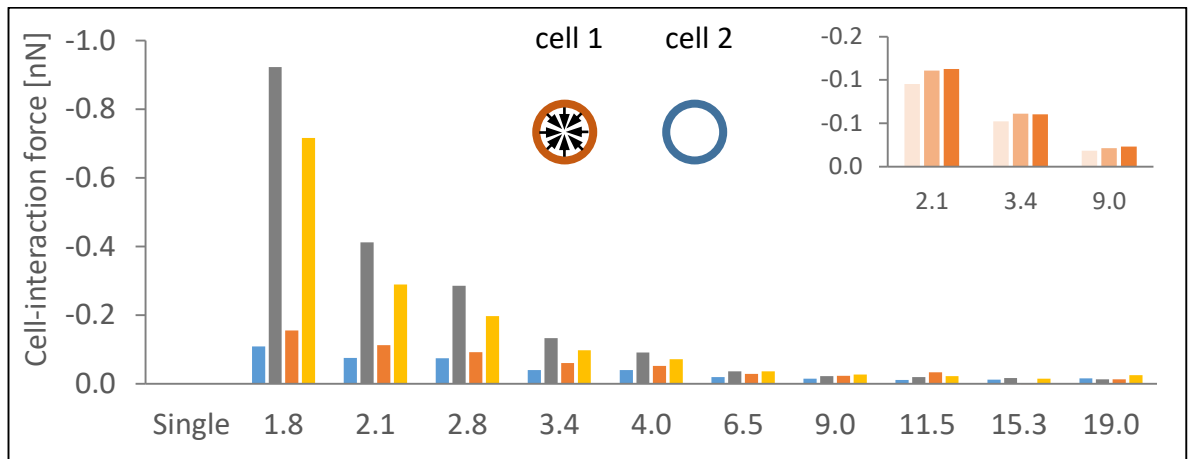

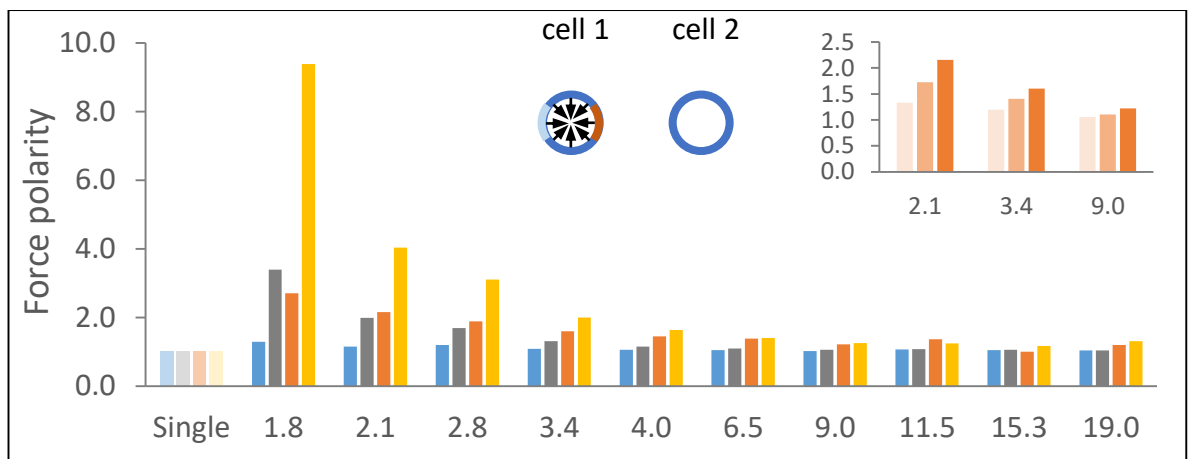

**Figure 7:** (a) Net cell-interaction force occurring on the cell boundary (as projected on the line connecting the cell centers), for 25% contraction. The interaction force is defined as positive for repulsion and negative for attraction. (b) The polarity ratio of the contraction force occurring on the cell boundary. Polarity ratio of 1.0 indicates no preferential orientation of loads toward the neighboring cell. Equivalent bar charts showing the net cell-interaction force and the contraction force polarity ratio for 10% and 50% contraction are contained as supplementary materials.

### 3.6. Effect of nonlinear elasticity on structural remodeling of the inter-cellular matrix

Structural remodeling of the ECM, including fiber densification and alignment, can influence the biological activity of cells, including morphology and migration (34), and is therefore of particular interest. Accordingly, we examined how the above-described preferred directionality of loads and their tendency to concentrate in the inter-cellular matrix influence fiber density and alignment. We found that the inter-cellular region of the matrix contained more fibers as a result of contraction. The level of increase in fiber density was considerably greater when the fibers buckled under compression, with strain stiffening under tension further contributing to this effect. The level of increase was greater for higher levels of cell contraction (Figure 8a, Figure S13a and Figure S14a). The fibers contained in the inter-cellular medium were also more aligned along the line connecting the cell centers, which is the direction of maximum tensile loading imposed by such cell contractions. The alignment was more considerable when the fibers buckled under compression and stiffened under tension, with buckling again showing a more considerable effect. Higher levels of cell contraction resulted in a more considerable alignment (Figure 8b, Figure S13b and Figure S14b). Similarly to the model outcomes discussed above, these trends – along with the changes in fiber density and orientation themselves – were more noticeable when the two neighboring cells were closer together. Overall, nonlinear elasticity of the fibers forming the ECM contributes to the increase in the density and alignment of the fibers contained in the inter-cellular region of the matrix, facilitating the formation of the 'bands' visible in our experimental findings (Figure 2) as well as in other biological contexts (3, 4, 34).

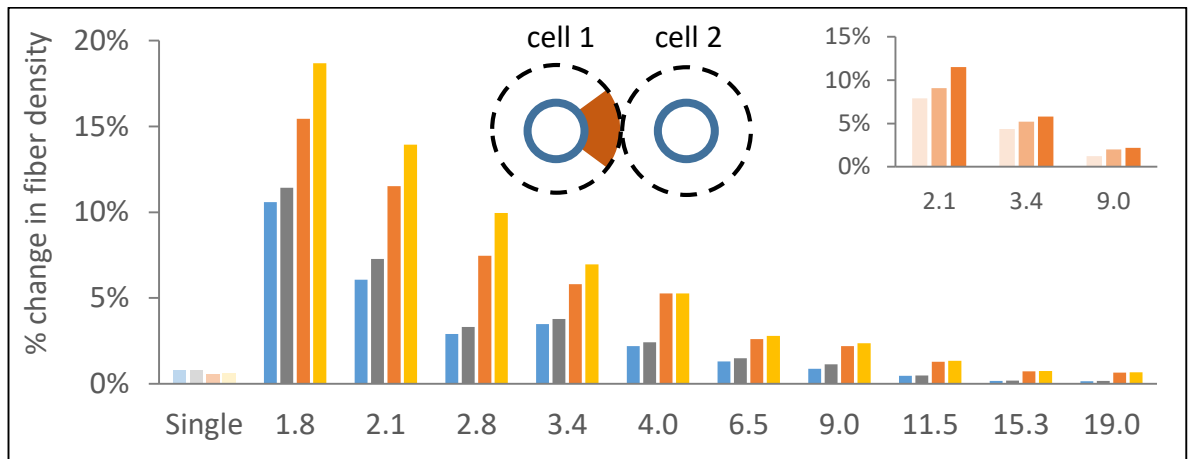

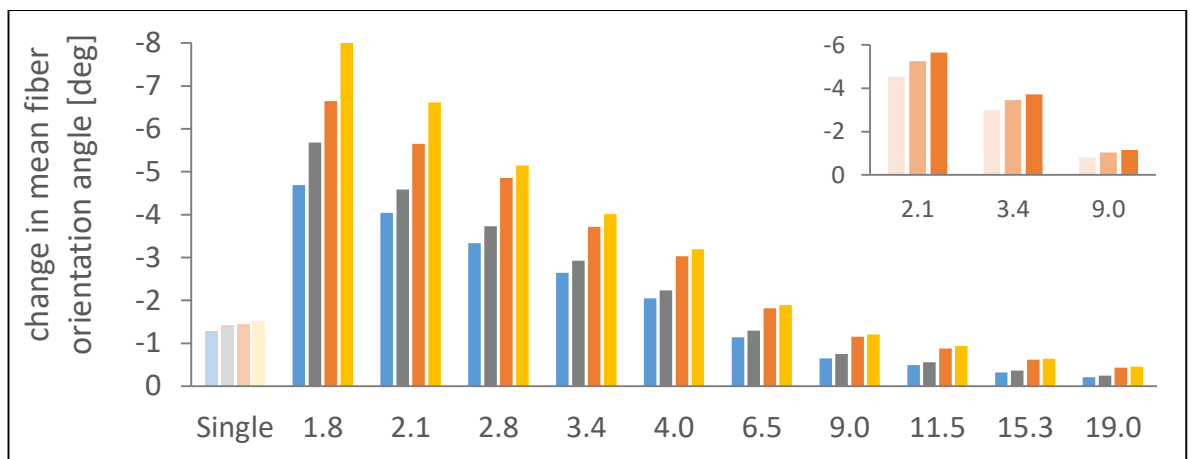

**Figure 8:** (a) Relative change in density of fiber segments contained in the inter-cellular medium, and (b) mean change in the angles of the orientation of the fibers contained in the inter-cellular medium (orange sector at the top right panel), as a result of 25% cell contraction. 0° is defined as the horizontal line pointing toward the center of the neighboring cell; accordingly, negative change indicates realignment of the fiber to point toward the center of neighboring cell (*i.e.* to be aligned more horizontally).

# 4. Discussion

The computational model presented in this study suggests that nonlinear elastic behavior of the fibers contained in the ECM – manifested in compression-microbuckling and tension-stiffening – results in the loads occurring in the fibers in the vicinity of contracting cells being more concentrated within the inter-cellular regions of the matrix and more directed toward neighboring contracting cells. Efficiency can be regarded as the quality of a process to occur with a minimum amount of wasted efforts; in more quantitative terms, it is the ratio of useful output to the total input. In the context of the current model, efficiency can be quantified as the fraction of cell-induced loads applied in the direction pointing toward the neighboring cell (and thus potentially result in mechanical signaling) to the total load induced by the cell contraction. Accordingly, the directionality-ratio outcomes presented herein (Sections 3.3 and 3.4) indicate that nonlinear elastic behavior of the individual fibers constituting biological gels contributes to the efficiency of the transfer of mechanical signals between neighboring cells, which may trigger biological response. The model outcomes further show that cells exert less energy when contracting within a matrix consisting of bucklable fibers, which is attributable to the smaller resistance of the matrix to deformations (Section 3.3); this further highlights the contribution of fiber buckling to efficient mechanical communication between distant cells via the ECM. While previous models have revealed the important contribution of elastic nonlinearity of biological gels to increasing the range of propagation of cell-contraction-induced loads through the ECM and specifically between neighboring cells (22, 26, 27, 29), our model is the first to demonstrate and quantify the direct effects of the mechanical behavior of the individual fibers on the directionality of mechanical loads and the efficiency of load transfer between cells.

The concentration of loads and the rearrangement of fibers within the inter-cellular medium (experimental: Section 3; computational: Section 3.6), which were observed in all model variants (though to different levels, depending on the modeled fiber elastic behavior, level of cell contraction and cell-to-cell distance), are ascribable to the simultaneous contraction of two cells that is practically applying tension to the inter-cellular band, while other regions of the network are not subjected to such elevated loading. The increased tension in this region of the ECM obliges the fibers to align, and also results in compression occurring perpendicularly – which is augmented by fiber compression-buckling – leading to fiber densification in the inter-cellular medium (the level of increase in fiber density within the inter-cellular region was found to be larger when fibers were bucklable; Section 3.6). This may induce 'compressive stiffening' of the inter-cellular ECM band (35, 36), particularly when the fibers are modeled as bucklable (36). The increased effective stiffness of this region of the matrix, which is subjected to elevated tension, can lead to the observed preferential orientation of loads toward the neighboring cell, which is more evident in fiber-buckling networks. The further increase in the directionality of loads (mainly SEDs) owing to tension-stiffening (Section 3.4) is attributable to the greater resistance of the strain-stiffened fibers (including these consisting the inter-cellular ECM band) to tensile deformation (*i.e.* stiffening).

While previous studies (15, 26, 27, 29) have captured the tendency of cell-contraction-induced loads to concentrate in the inter-cellular regions of the matrix, our study provides a systematic quantitative evaluation of the transfer of mechanical signals between neighboring cells, and the dependence of which on the nonlinear elastic mechanical properties of the matrix fibers and the distance between the cells. For example, Notbohm *et al.* (26) and Abhilash *et al.* (29) showed that cell-contraction-induced tensile strains occur almost entirely in the band between the two cells, which was ascribed to fiber microbuckling. Ma *et al.* (27) further suggested that 24 hours post cell seeding in collagen gel, the preferential orientation of cell-contraction-induced stresses toward neighboring cells is facilitated by increased fiber alignment. In our simulations we show that preferential orientation of cell-contraction-induced loads toward neighboring cells can occur without *a priori* fiber alignment, but both the load orientation and fiber alignment occurred simultaneously with cell contraction.

The increased fiber density and alignment within the inter-cellular band – which was demonstrated in our simulations to be most considerable when the individual fibers contained in the ECM were modeled as of nonlinear mechanical behavior (Section 3.6) – is corroborated by our experimental findings (Section 3), and is in agreement with the cell-contraction-induced fiber densification and alignment previously demonstrated both *in vitro* (1, 3, 26, 27, 32) and *in silico* (16, 28–30, 32). Previous biological experiments have further suggested that the formation of inter-cellular bands of dense and aligned fibers regulate cell function in various biological processes, including directing the growth of capillary sprouts originated in spheroids in collagen gels (3), controlling cancer invasion in mechanically interacting acini (4, 32) and directing fibroblast patterning (20). This diversity of biological contexts implies that long-range inter-cellular mechanical interactions via fibrous gels has a universal role in cell and tissue function. Our model highlights the important role played by the nonlinear elastic behavior of the ECM fibers in the formation of these inter-cellular bands.

Our FE model allowed us to directly infer the interaction forces acting between neighboring contracting cells and examine their dependency on the cell-to-cell distance, as opposed to previous analytical models that analyzed the interaction based on energy considerations (37, 38). In our simulations, we found that neighboring contracting cells always attract each other up to a cell-cell distance of 6-7 cell diameters, with the attraction forces being larger when the cells are closer together. Previous experimental study found that endothelial cells cultured on linear synthetic gels apply both repulsive and attractive forces to each other (21). The different behaviour may be due to the fibrillar structure of the ECM, which is absence in synthetic gels.

Our simulations predict inter-cellular mechanical signal transmission to occur up to a distance of 15 cell-diameters apart. Specifically, we observe structural changes within the inter-cellular matrix to occur up to a distance of 15.3 cell-diameters (alignment and densification), tensile and SED's directionality ratios predict preferential load transmission to occur up to 9 cell-diameters, and force interaction predicts sensing up to 6.5 cell-diameters. The influence of fiber tension-stiffening on inter-cellular loads is notable up to a cell-to-cell distance of only 4 cell diameter (Section 3.4), while the effect of compression-buckling spans an over-twice longer distance (9-15 cell diameters). This is supported by the recent analytical predications of Xu *et al.* (22), which imply that tension-stiffening dominates the propagation of matrix displacements in the near vicinity of the contracting cell, whereas compression-buckling dictates the distribution of displacements in the further regions of the matrix. The range of inter-cellular mechanical force transmission predicted by our simulations (approximately 12 cell diameters or 175 μm when assuming that the fibers are bucklable; Section 3.3) is larger than that of Humphries *et al.* (15) (5 cell diameters), but similar to that of Wang *et al.* (30); it is also somewhat larger and considerably smaller than predicted in the experimental studies by Ma *et al.* (27) (120 μm) and Winer *et al.* (20) (500 μm), respectively.

Like in many previous models (15, 16, 22, 25, 33), cells were modeled here as circles. This simplified representation captures the morphology of cells in the first hours after seeding in biological gels (Section 3, Figure 2; (27)). This simplification is in line with the purpose of our model, designed to allow investigation of the mechanical signaling occurring between neighboring contracting cells embedded in biological gels shortly after seeding, prior to cell morphogenesis, and while being isolated from any modifications potentially occurring in the ECM at later stages. Similarly, we assumed uniformly distributed (*i.e.* isotropic) contractile displacements of the cell membrane. While this simplification is likely to considerably affect the simulation-predicted loads occurring in the very near vicinity of the contracting cell, it has less influence farther away from the cell (27). Furthermore, like in most previous models, prior to cell contraction the fiber segments were assumed to be unstressed, which is again in line with the aim of this simulation – capturing cell-ECM and inter-cellular mechanical interactions at 'time zero'. Overall, our computational model mimics the biological scenario of the early stages of cell-ECM and cell-cell interactions and predicts the ability of cells to initiate mechanical cues and communicate through mechanical pathways shortly after their seeding. These early mechanical cues can guide cell growth and expansion toward nearby cells, as previously described in fibroblasts spreading toward each other already at 5-7 hours post seeding (26).

Our model has notable limitations that should be considered when interpreting the results. First, similarly to most previous models (15, 16, 27, 29), the model we developed to simulate an inherently 3D system, is 2D. Such simplification, however, possesses a considerable computational advantage, which allowed us to run a large number of model variants in order to explore the effects of several factors on the outcome measures of interest, while still capturing all aspects of network mechanics, including nonlinear mechanical loading response in the macroscale level and loading-induced fiber realignment (29). Additionally, the mechanical properties assigned to the individual fibers were not directly derived from the literature, but from the simulated macroscale properties of the network, which were juxtaposed against previous experimental findings of the bulk response of collagen gels to uniaxial loading (Supplementary Materials). This is because most of our understanding of the mechanical behavior of fibrous gels comes from bulk measurements, while data on the elastic behavior of the individual fibers contained in the gels are sparse and inconsistent (16). Furthermore, we modeled the fiber segments as truss elements, which encompasses the assumption that these are subjected to uniaxial tension and compression (without bending) and are able to rotated freely one with respect to the other (Section 2.2.4). Some of the previous models (16, 25, 27, 29, 30, 33), however, did account for bending potentially occurring in the fibers and/or between two fibers meeting at a crosslink. Nevertheless, a preliminary study we conducted showed that using beam elements had affected the model outcomes only slightly, which concurs with previous findings (15, 16, 26, 33, 39). Specifically, Humphries *et al.* (15) and Heussinger and Frey (39) found that even when bending dominated the stretching energy of the loaded fibers, the forces acting in them were still predominantly axial, while Abhilash *et al.* (29) further claimed that the loading response of 2D models is less dominated by bending of the network segments. Considering these and other potential limitations of the model, it is recommended that the data presented in this study be interpreted mostly as trends of effects, rather than as absolute values (40). Particularly, the objective of the study was not to reproduce specific numerical results reached through laboratory experiments, but to compare trends of effects of the mechanical behavior of the ECM fibers on the mechanical signaling between neighboring cells, and we do not expect the limitations discussed above to affect such comparative findings.

To conclude, the FE model presented herein predicts that nonlinear elastic behavior of the individual fibers constituting the ECM contributes to a highly directional and efficient inter-cellular mechanical signal transmission. Such mechanism regulating long-range cell-cell communication can elucidate previous experimental observations of biological processes involving inter-cellular mechanical signaling; it further highlights the importance of utilizing fibrous biological gels for facilitating long-range inter-cellular communication, in comparison with linear-elastic synthetic gels (such as polyethylene glycol and polyacrylamide gels), which are commonly used for cell culturing (41). Major challenges for future work include a more realistic description of the physiological scenario, including: 3D modeling; a more accurate representation of the network architecture in terms of connectivity and fiber arrangement; dynamic modeling accounting for the time-dependent cell morphology and loading, as well as the viscoelastic and plastic nature of ECM.

**Acknowledgments**: We thank Prof. Samuel Safran and Prof. Yair Shokef for their recommendations while preparing the manuscript. This study was supported by the Israel Science Foundation number 1474/16 and the Israel Science Foundation- I-CORE number 1902/12.

# Supplementary Materials

**Derivation and validation of the mechanical properties assigned to the network fibers**

The mechanical properties assigned to the fibers contained in the ECM were derived from the simulated macroscale properties of the network, which were juxtaposed with previous experimental findings of the bulk response of collagen gels to uniaxial loading. In detail, a network consisting of truss elements of diameter of 200 nm (which is within the range reported by (17) for the diameter of collagen fibers), and of rectangular shape (length: 535 μm, width: 229 μm), was created as described in Section 2.2.1. The individual fibers were modeled as demonstrating all types of mechanical behavior listed in Section 2.2.2, while the reference elastic modulus ($E_{ref}$) was initially set at a random value. The bottom edge of this virtual specimen was fixed for all translations and rotations. Uniaxial tension was introduced by applying maximum displacement of 100-μm to the top edge of the specimen. The nominal strain applied to the specimen was continuously calculated by dividing the length change of the rectangle in the vertical axis (*i.e.*, the vertical displacement occurring at the top edge) by the reference, undeformed length (*i.e.*, the initial length of the rectangle). The nominal stress applied to the specimen was calculated as the sum of the vertical components of all reaction forces occurring at the upper edge of the rectangle, divided by the axial cross-sectional area of the specimen (the width of the rectangle multiplied by the 'depth' of the specimen, *i.e.* the fiber diameter of 200 nm). The calculated stresses were plotted against the calculated strains, and the resulted curve, which demonstrated the macroscale stiffness of the simulated fibrous material, was juxtaposed with the reported mechanical behavior of collagen gel 2.4 mg/ml subjected to uniaxial tension, as measured using a rheometer (9) (Figure S1). The value of $E_{ref}$ assigned to the individual fibers was iteratively adjusted until reaching satisfactory resemblance between the curves (a process similar to that described by (1)), and was ultimately set at 11.5 kPa.

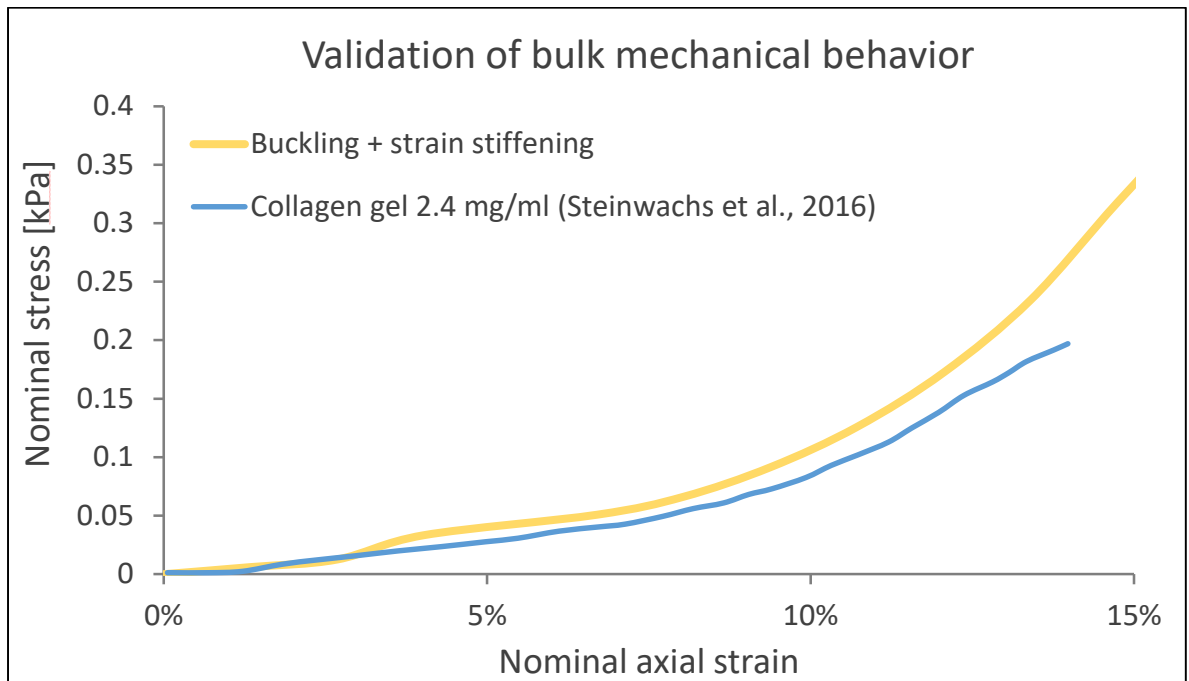

**Figure S1:** Stress-strain curve demonstrating the bulk mechanical behavior of the simulated fibrous network (when assuming fiber diameter of 200 nm, tension-stiffening of the individual fibers and $E_{ref}$=11.5 kPa) when subjected to uniaxial tension, juxtaposed with an equivalent curve derived experimentally for collagen gel 2.4 mg/ml (9).

The stress-strain relationships demonstrated by the simulated bulk material, as derived from the aforementioned analysis when assuming each of the four types of fiber mechanical behavior listed in Section 2.2.2 and implementing the aforementioned value of $E_{ref}$, are shown in Figure S2a. When assuming tension-stiffening behavior of the individual fibers (material model iii in Section 2.2.2), the bulk materials was the stiffest, followed by fibers demonstrating both tension-stiffening and compression-buckling (material model iv). In other words, the elevated resistance of the fibers contained in the network to tension resulted in an increased tension-stiffness of the bulk material. When assuming fiber compression-buckling alone (material model ii), the bulk material was slightly softer than when assuming linear elasticity (material model i in Section 2.2.2) (Figure S2a). This is attributable to the decreased resistance of the fibers subjected to compression loading (particularly those aligned horizontally, *i.e.* along the width of the specimen) to such compression.

The Poisson's ratio of the simulated material was also calculated when assuming all types of fiber mechanical behavior listed in Section 2.2.2, by dividing the nominal strain along the width of the rectangular specimen (transverse strain) by the nominal strain along the length of the material (axial strain). When assuming linear-elastic behavior of the individual fibers, Poisson's ratio was nearly constant at 0.4-0.5 (Figure S2b). When modeling the fibers as bucklable, increasing axial strain resulted in a gradual increase in the Poisson's ratio, which is in agreement with previous findings (26). Tension-stiffening of the individual fibers resulted in the Poisson's ratio of the simulated bulk material exceeding 1 (Figure S2b), which is in agreement with a previous computational model (32).

Pure shear was also exerted to the virtual specimen by applying horizontal displacement to the top edge of the rectangle. The engineering shear strain applied to the specimen was continuously calculated as the change in angle between the horizontal and vertical edges of the rectangle. The engineering shear stress was calculated as the sum of the horizontal components of all reaction forces occurring at the upper edge of the rectangle, divided by the axial cross-sectional area of the specimen. The shear modulus of the bulk material was subsequently estimated as the fraction of the shear stress to the shear strain applied to the specimen. The shear modulus was then plotted against the engineering shear strain (Figure S2c). Tension-stiffening of the individual fibers resulted in the shear modulus increasing with the shear strain. Fiber compression-buckling resulted in the bulk shear modulus being considerably smaller (Figure S2c), which is attributable to a decreased resistance of the fibers to compression.

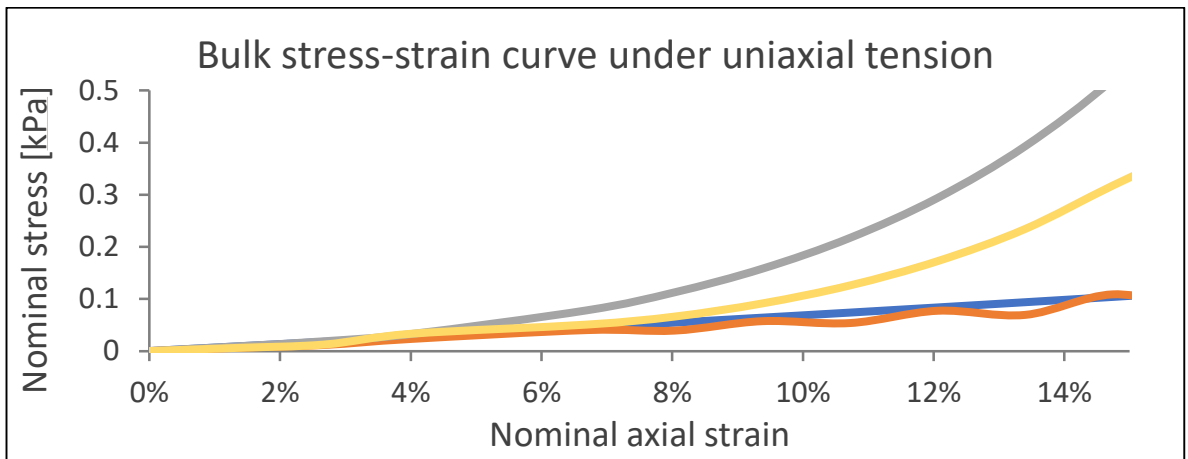

(a)

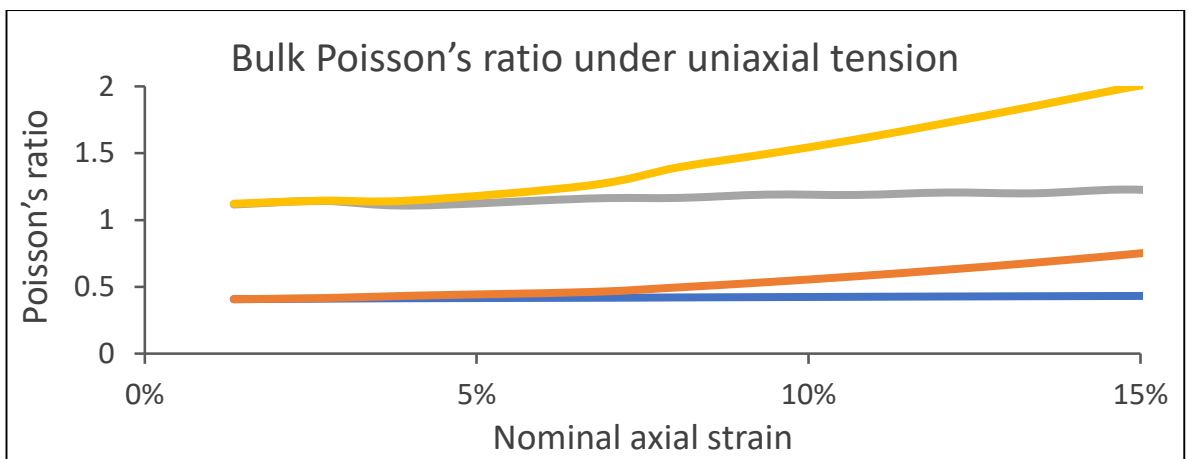

(b)

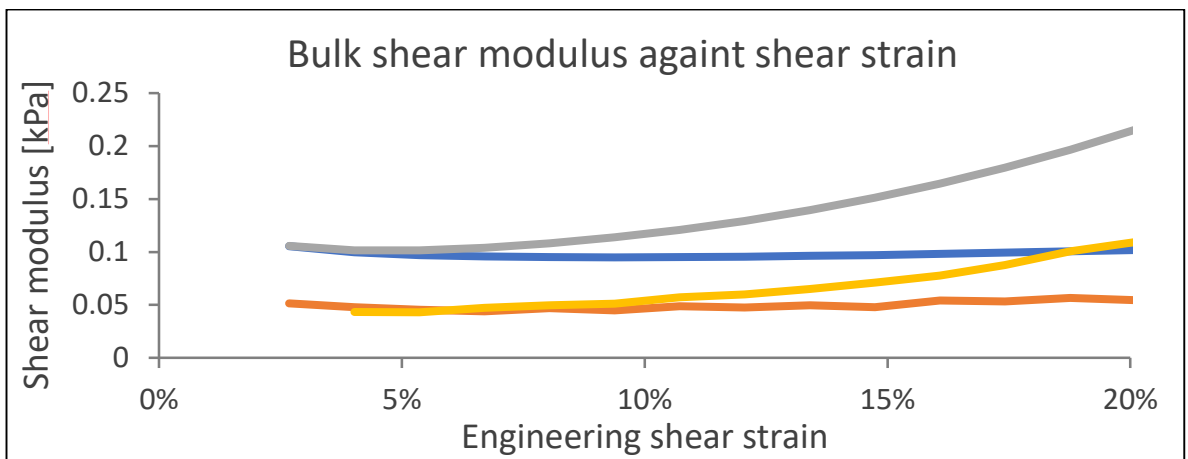

(c)

**Figure S2:** The bulk mechanical behavior of the simulated fibrous network (when assuming fiber diameter of 200 nm and $E_{ref}$=11.5 kPa) for four types of mechanical behavior of the individual fibers: (a) stress-strain relationship under uniaxial tension; (b) Poisson's ratio under uniaxial tension; (c) shear modulus against engineering shear strain.

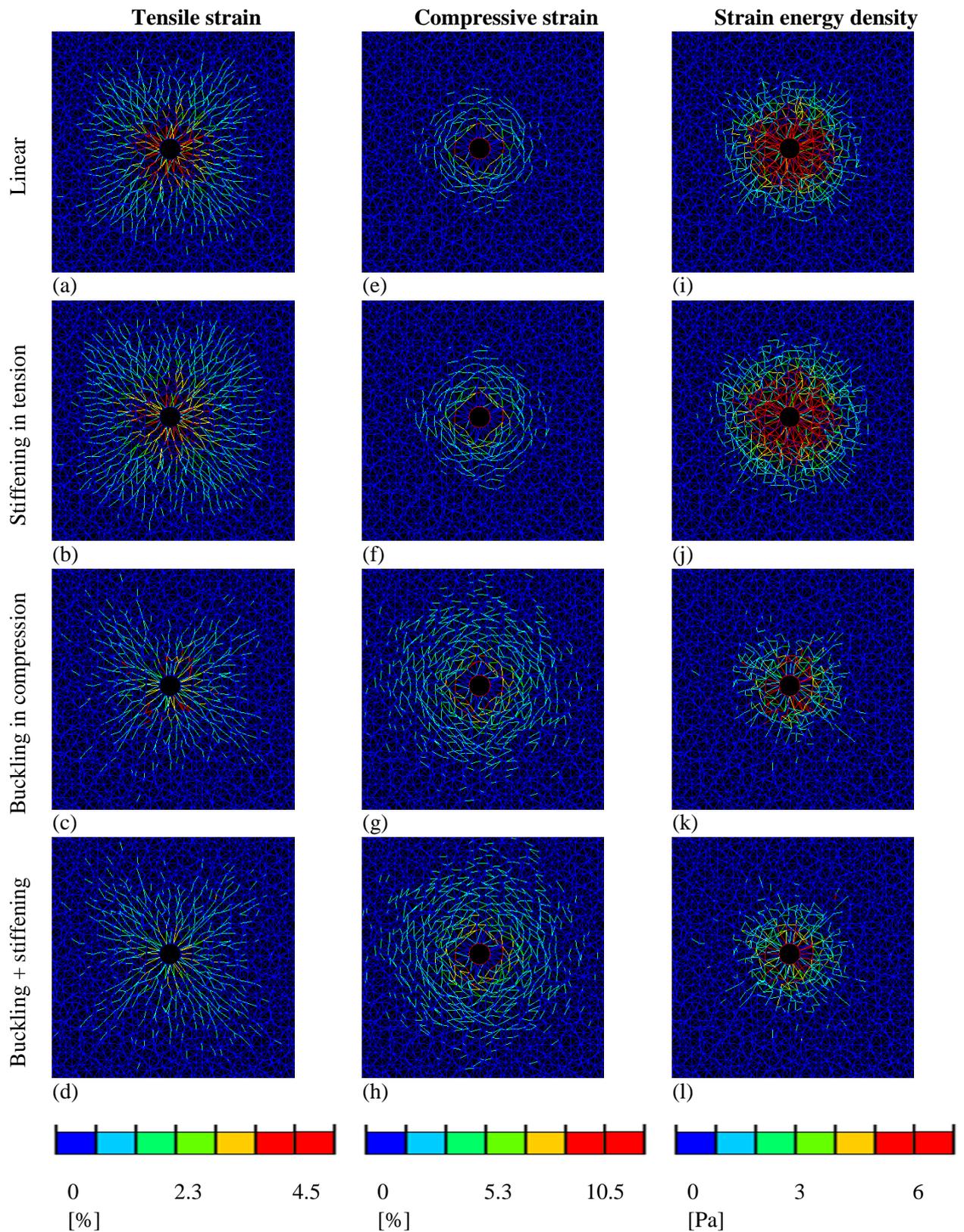

**Figure S3:** Contour plots showing the tensile strains (left column), compressive strains (middle column) and SEDs (right column) occurring in the fiber segments within the vicinity of a single, isolated contracting cell, for 25% contraction. Plots were produced for all four material models used to simulate the mechanical behavior of the individual fibers (Figure 1c).

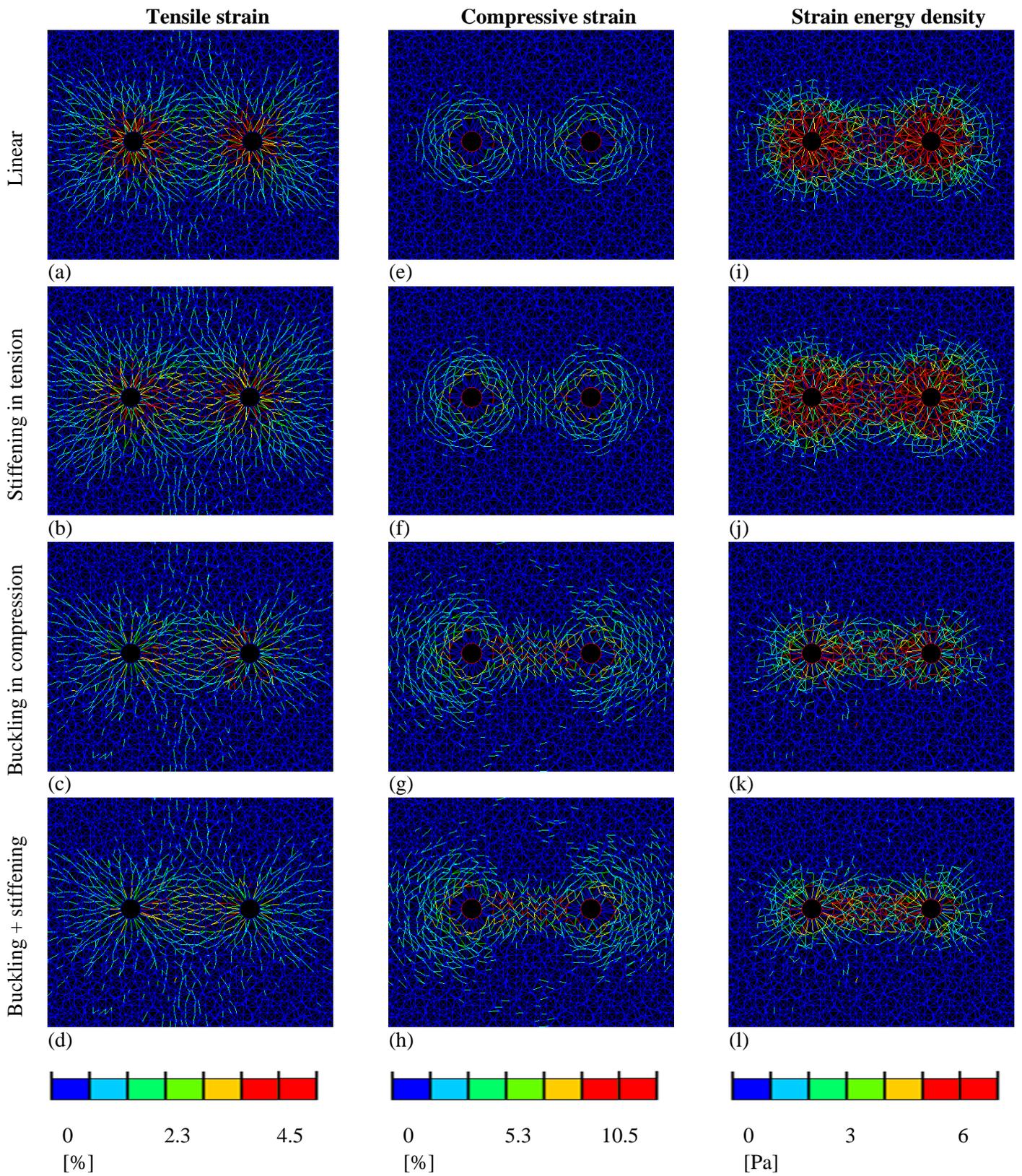

**Figure S4:** Contour plots showing the tensile strains (left column), compressive strains (middle column) and SEDs (right column) occurring in the fiber segments within the vicinity of two neighboring contracting cells (here the cell-to-cell distance is equivalent to 3.4 cell diameters as an example), for 25% contraction. Plots were produced for all four material models used to simulate the mechanical behavior of the individual fibers (Figure 1c).

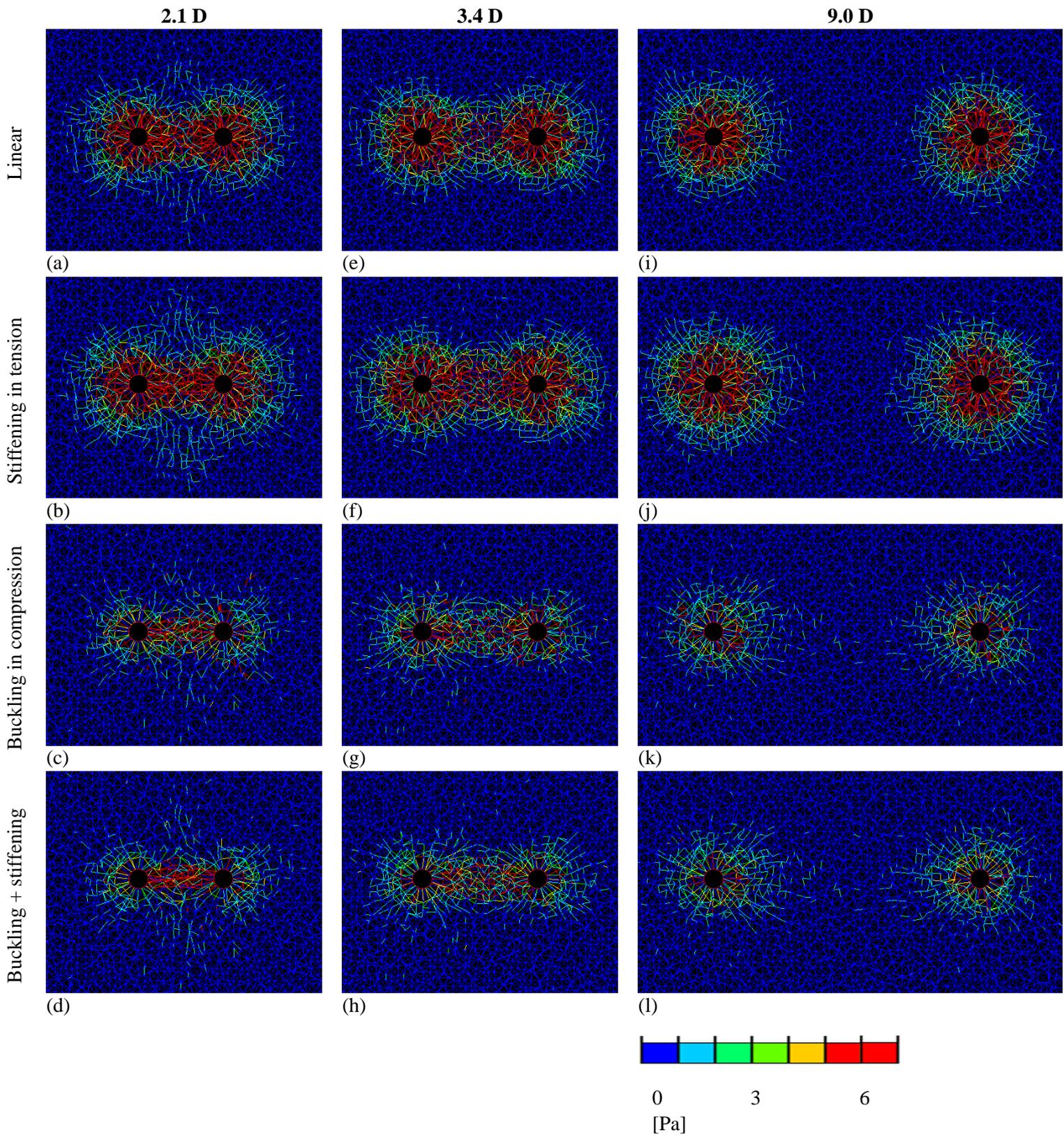

**Figure S5:** Contour plots showing the SEDs occurring in the ECM fiber segments surrounding two contracting cells, for 25% contraction. Cell-to-cell distances shown here are equivalent to 2.1 (left column), 3.4 (middle column) and 9.0 (right column) cell diameters. Plots were produced for all four material models used to simulate the mechanical behavior of the individual fibers (Figure 1c).

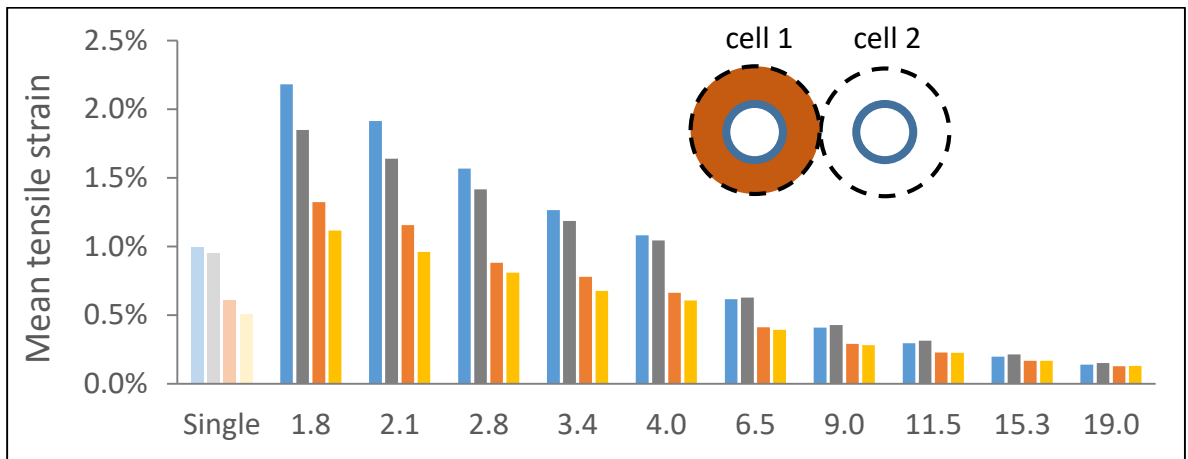

(a)

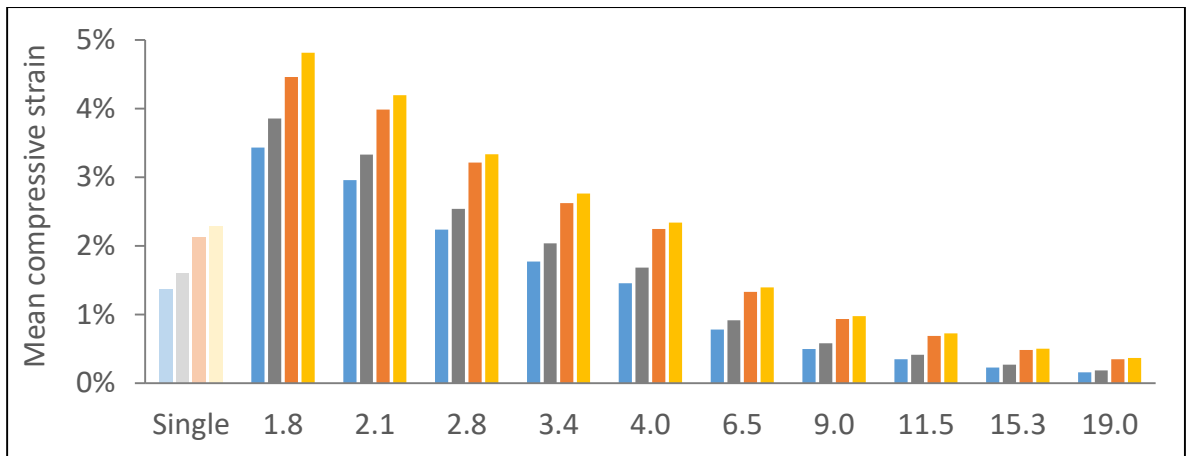

(b)

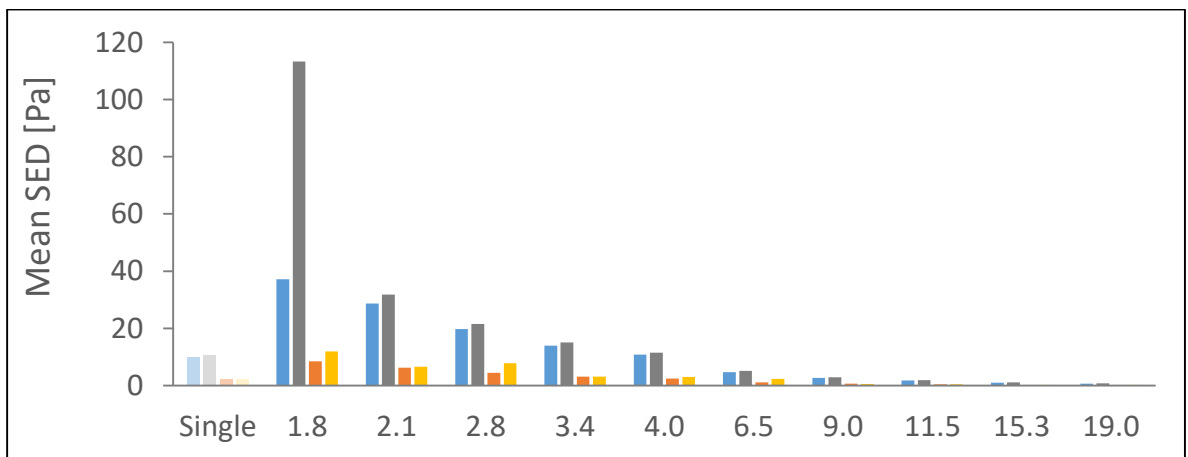

(c)

— Linear　— Strain stiffening　— Buckling　— Buckling + strain stiffening

**Figure S6:** Mean tensile (logarithmic) strain (a), compressive (logarithmic) strain (b) and SED (c), occurring within a disc surrounding an individual cell, of radius equals to half of the cell-to-cell distance, for 25% contraction. The model variants shown include several cell-to-cell distances (in terms of cell diameter, D), as well as the equivalent single-cell model variants, and four of the material models used to simulate the mechanical behavior of the individual fibers (Figure 1c).

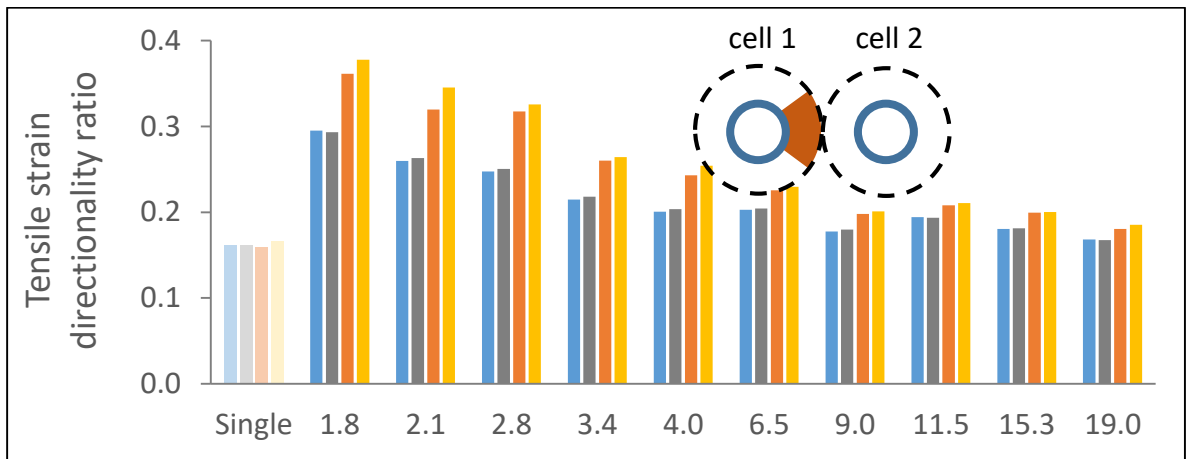

(a)

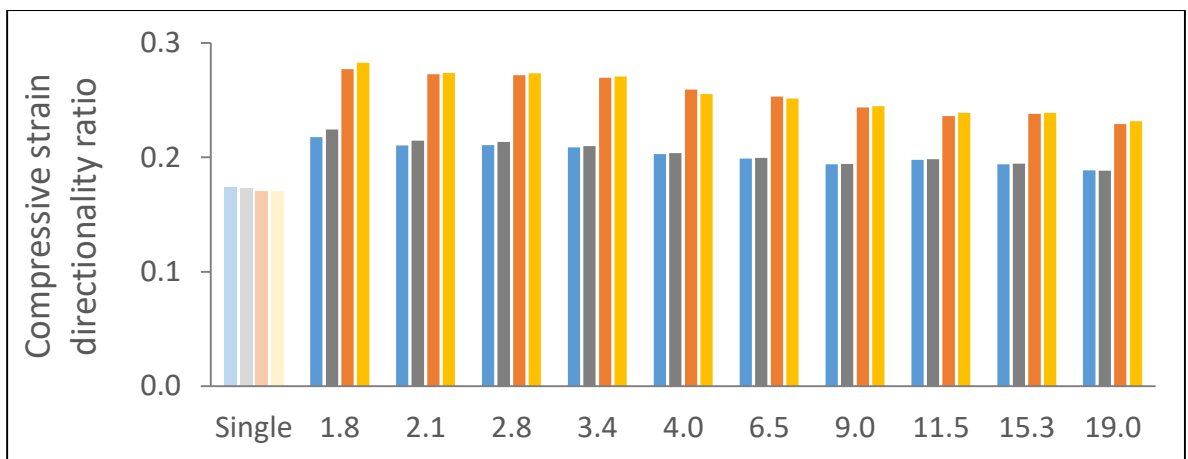

(b)

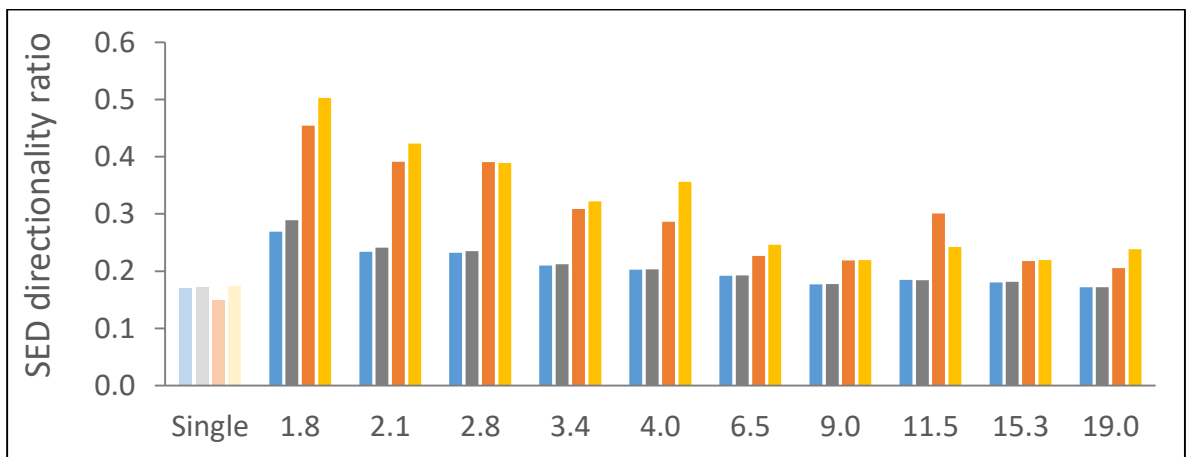

(c)

— Linear   — Strain stiffening   — Buckling   — Buckling + strain stiffening

**Figure S7:** Directionality ratios for tensile strains (a), compressive strains (b) and SEDs (c), occurring within a disc surrounding an individual cell, of radius equals to half of the cell-to-cell distance, for 10% contraction.

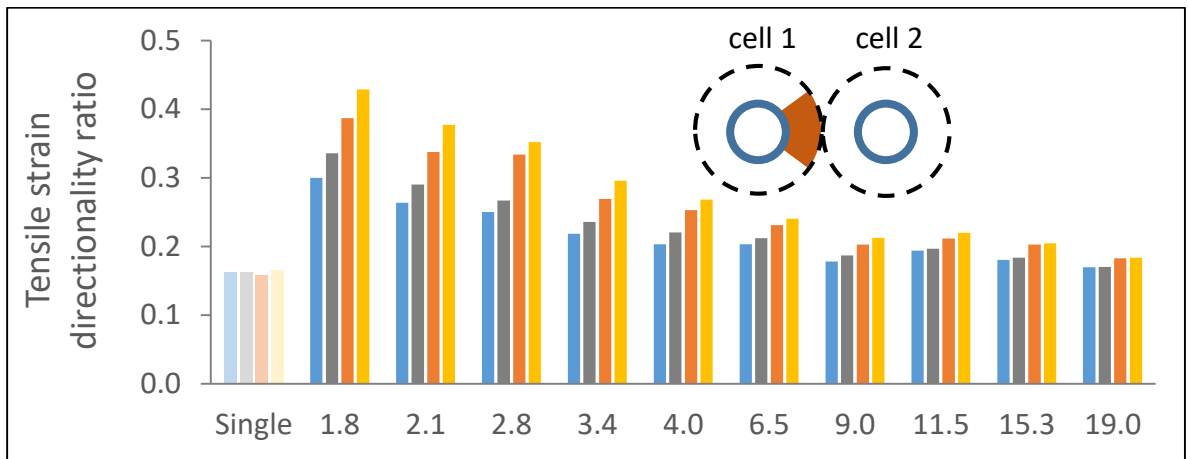

(a)

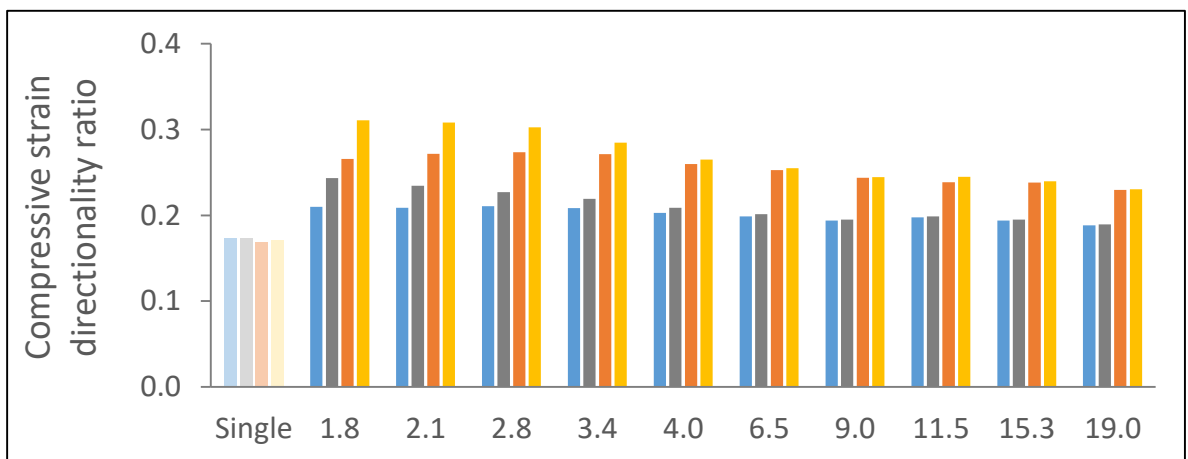

(b)

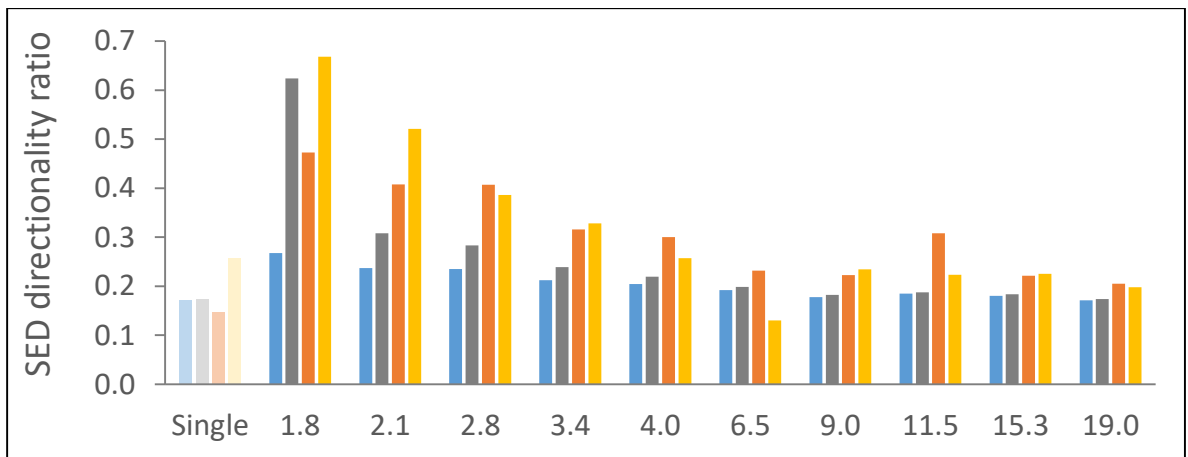

(c)

— Linear  — Strain stiffening  — Buckling  — Buckling + strain stiffening

**Figure S8:** Directionality ratios for tensile strains (a), compressive strains (b) and SEDs (c), occurring within a disc surrounding an individual cell, of radius equals to half of the cell-to-cell distance, for 25% contraction.

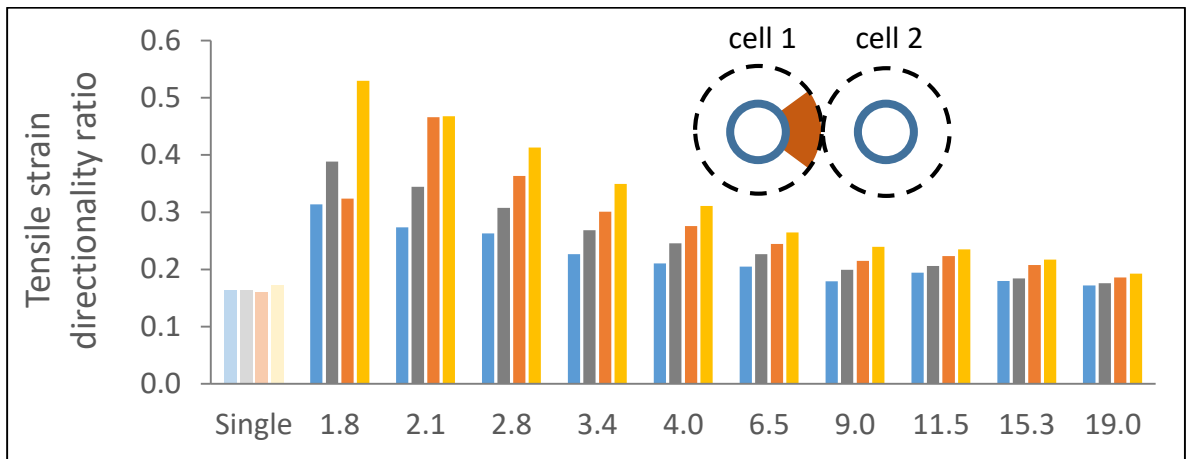

(a)

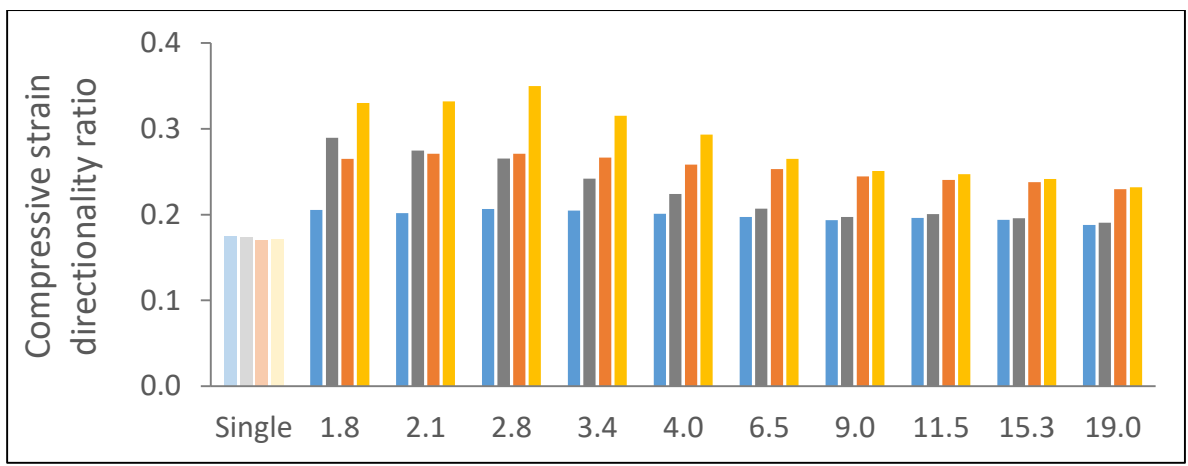

(b)

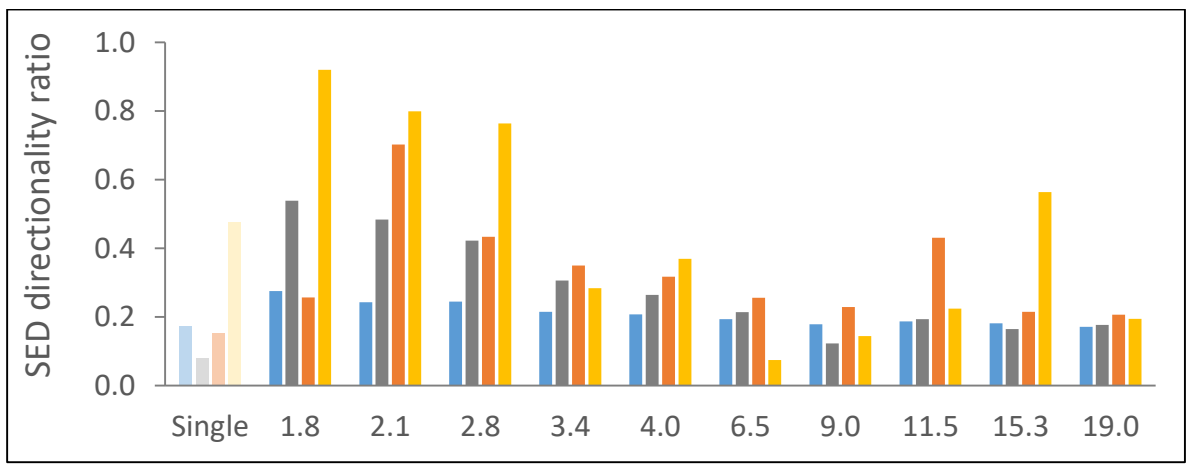

(c)

— Linear   — Strain stiffening   — Buckling   — Buckling + strain stiffening

**Figure S9:** Directionality ratios for tensile strains (a), compressive strains (b) and SEDs (c), occurring within a disc surrounding an individual cell, of radius equals to half of the cell-to-cell distance, for 50% contraction.

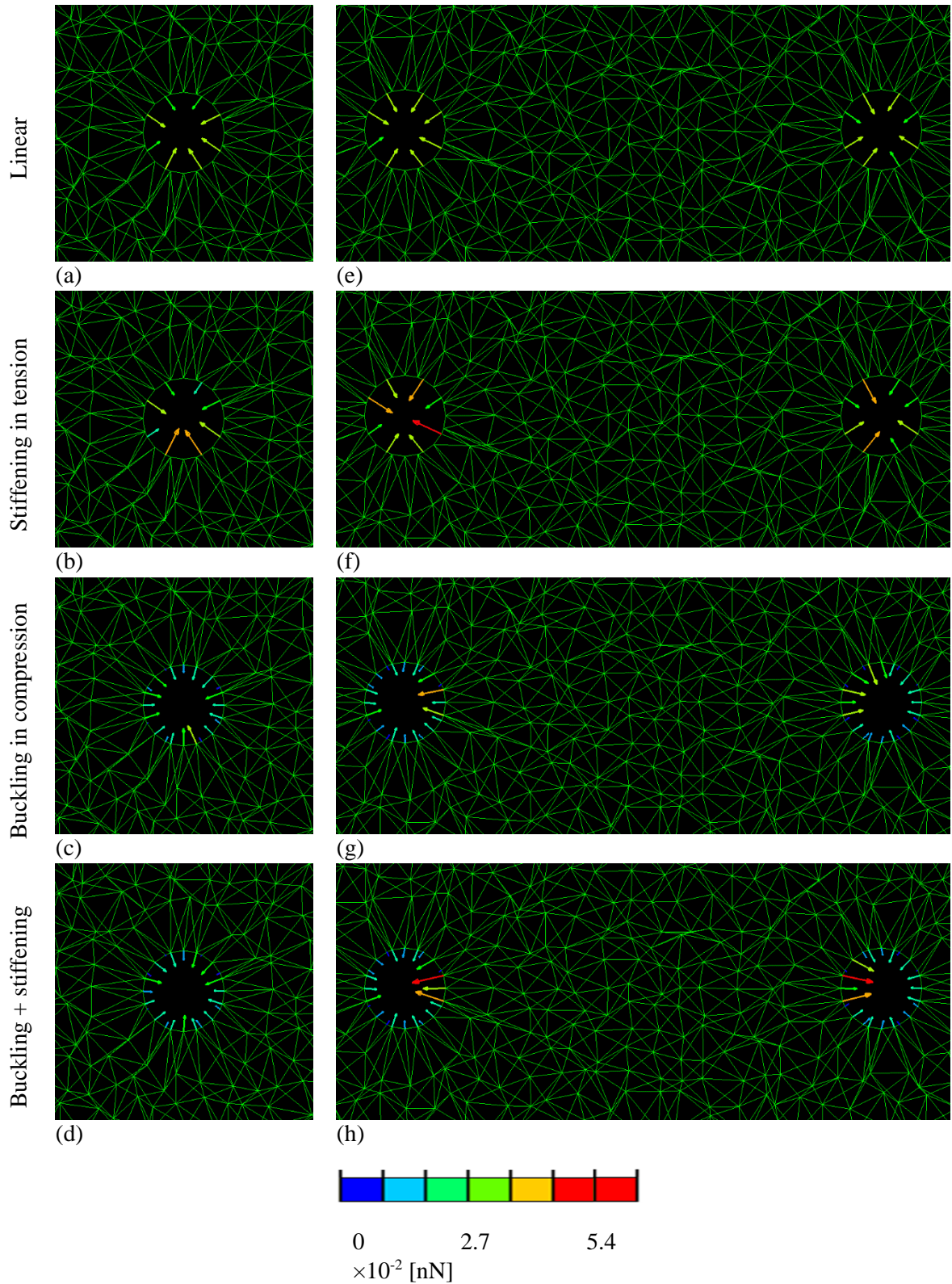

**Figure S10:** Contour plots showing the reaction forces occurring on the cell boundaries for a single (left column) or two (right column; here distance between the neighboring cells is equivalent to 3.4 cell diameters as an example) contracting cells, for 25% contraction. Plots were produced for all four material models used to simulate the mechanical behavior of the individual fibers (Figure 1c).

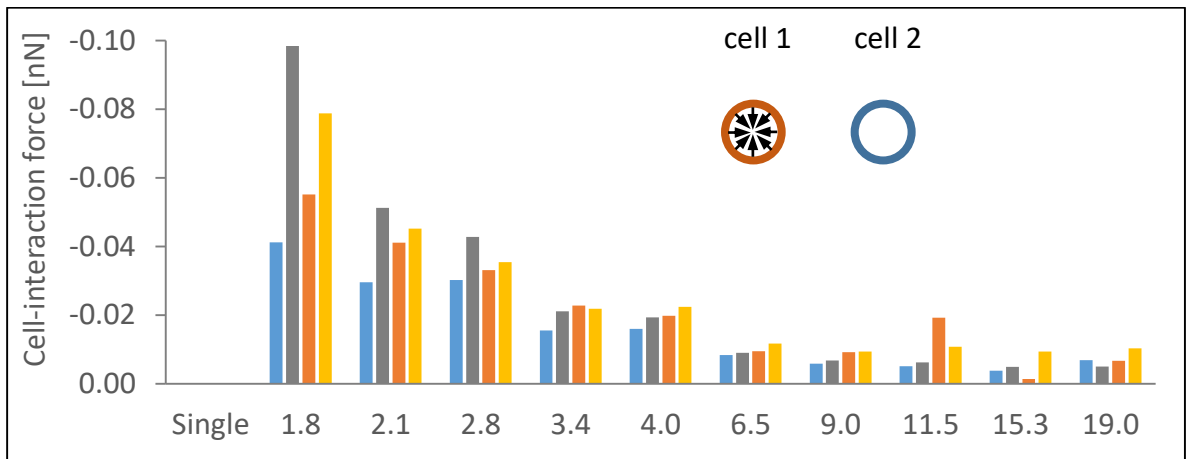

(a)

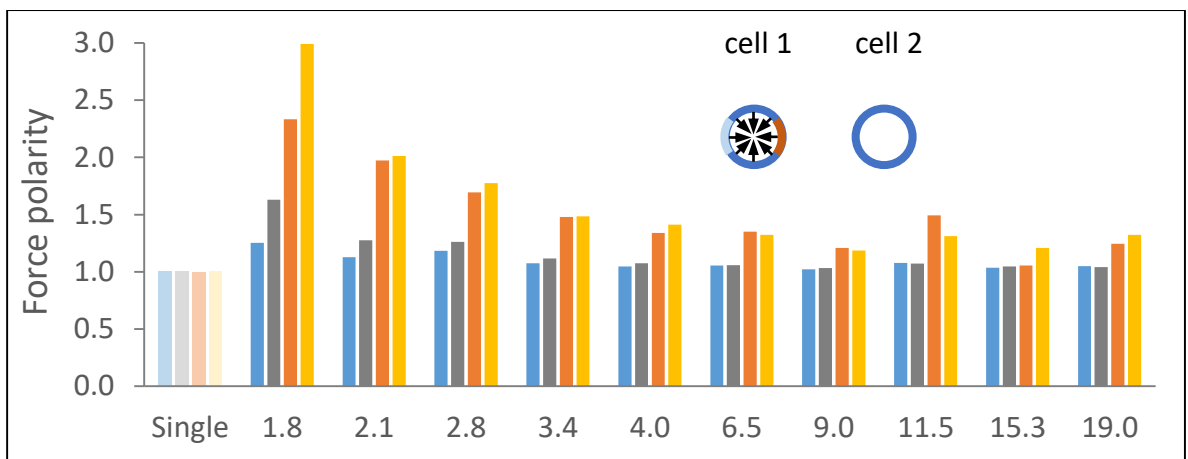

(b)

**Figure S11:** (a) Net cell-interaction force occurring on the cell boundary (as projected on the line connecting the cell centers), for 10% contraction. (b) The polarity ratio of the contraction force occurring on the cell boundary.

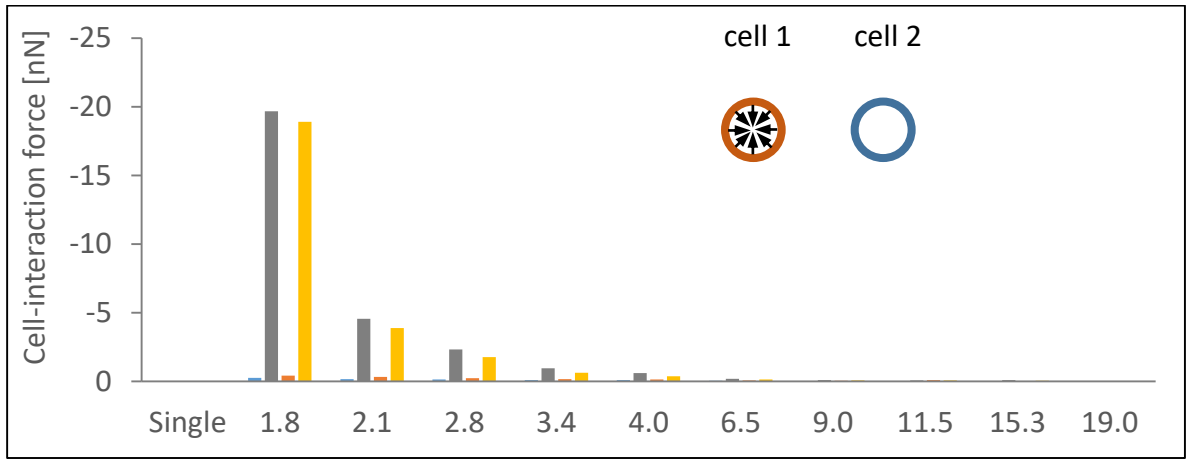

(a)

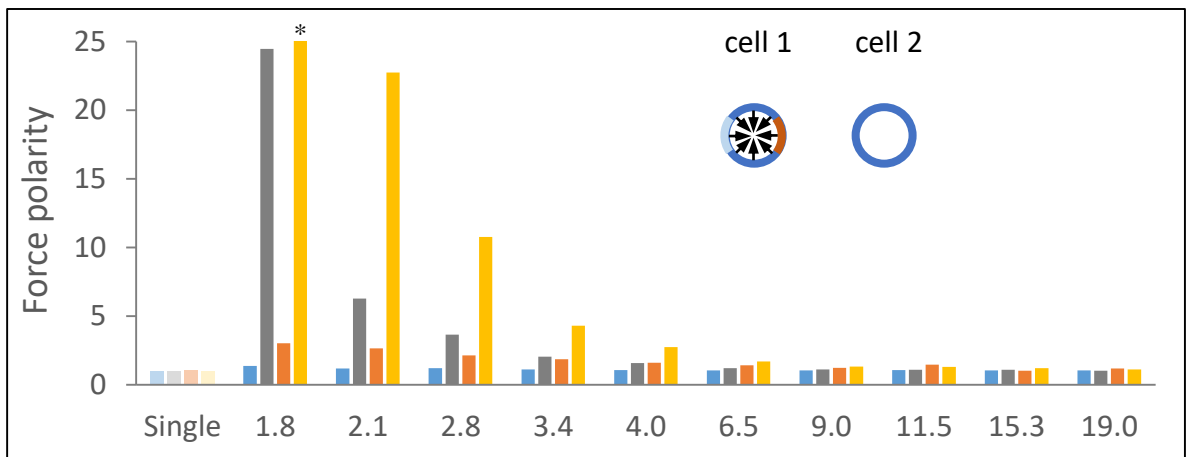

(b)

**Figure S12:** (a) Net cell-interaction force occurring on the cell boundary (as projected on the line connecting the cell centers), for 50% contraction. (b) The polarity ratio of the contraction force occurring on the cell boundary.
\* The value should actually be approximately 109, but is not shown on the graph.

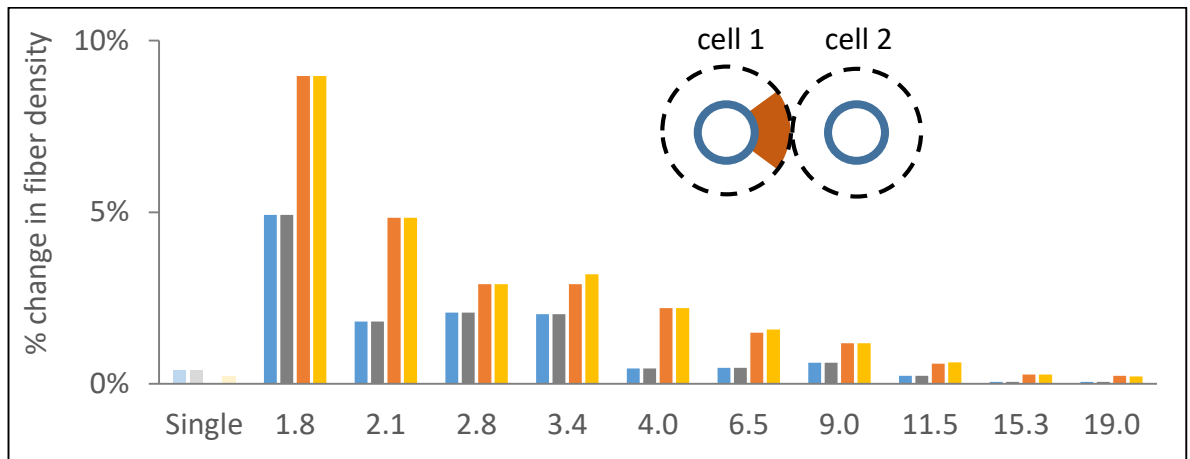

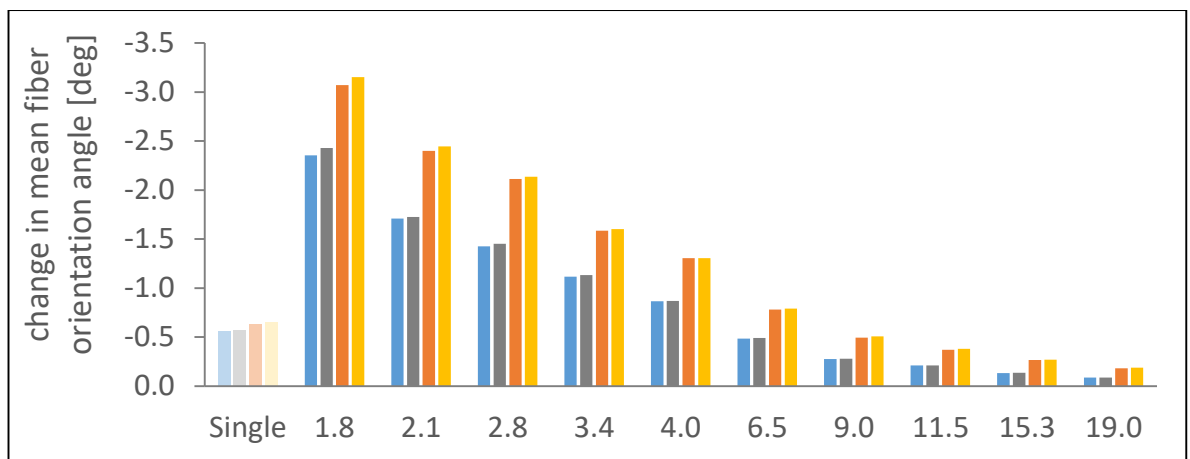

**Figure S13:** (a) Relative change in density of fiber segments contained in the inter-cellular medium, and (b) mean change in the angles of the orientation of the fibers contained in the inter-cellular medium (orange sector at the top right panel), as a result of 10% cell contraction.

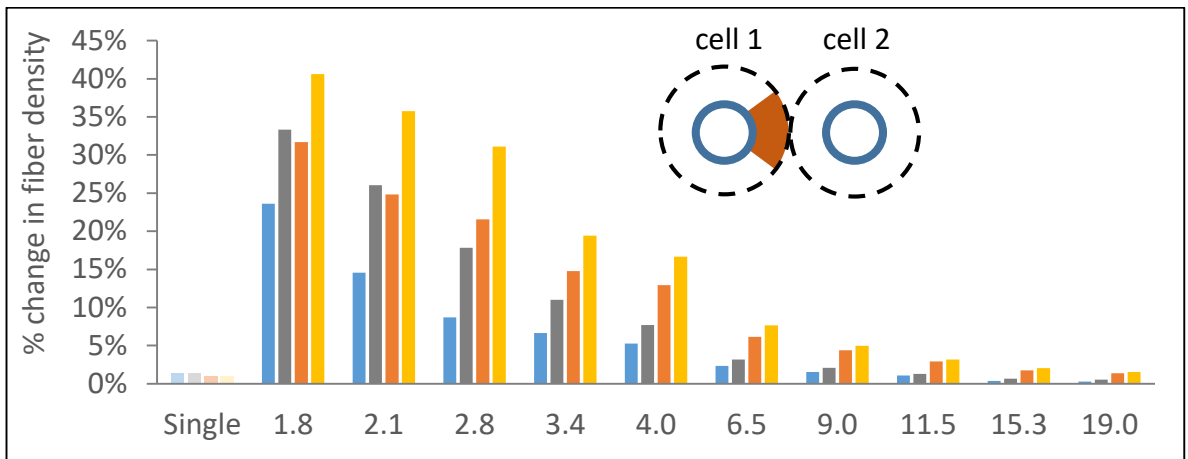
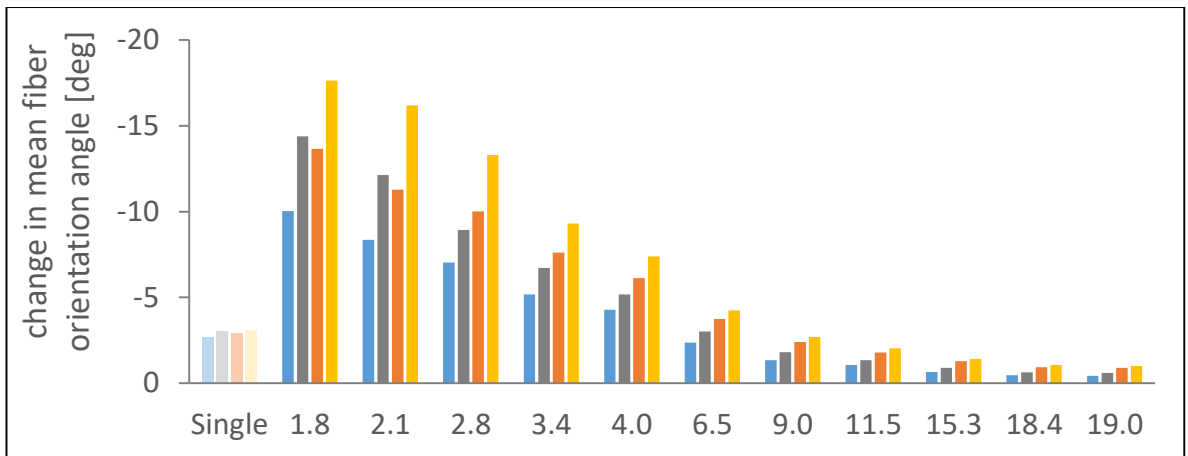

**Figure S14:** (a) Relative change in density of fiber segments contained in the inter-cellular medium, and (b) mean change in the angles of the orientation of the fibers contained in the inter-cellular medium (orange sector at the top right panel), as a result of 50% cell contraction.